\documentclass[10pt, twocolumns]{IEEEtran}
\usepackage{amsmath}
\usepackage{comment}
\usepackage{amssymb} 
\usepackage{graphicx} 
\usepackage{cite}
\usepackage{algorithm}
\usepackage{algorithmic}
\usepackage{array}
\usepackage{float}
\usepackage{multirow} 
\usepackage{booktabs}
\usepackage[font=small]{caption}
\usepackage[export]{adjustbox}
\usepackage[caption=false,font=footnotesize]{subfig}
\usepackage[table,dvipsnames]{xcolor}

\newcommand{\RN}[1]{\textup{\uppercase\expandafter{\romannumeral#1}}}

\def\BibTeX{{\rm B\kern-.05em{\sc i\kern-.025em b}\kern-.08em
    T\kern-.1667em\lower.7ex\hbox{E}\kern-.125emX}}

\title{Echo-Side Integrated Sensing and Communication via Space-Time Reconfigurable Intelligent Surfaces}
\author{Marouan~Mizmizi,~\IEEEmembership{Member,~IEEE,}
        Stefano~Tebaldini,~\IEEEmembership{Senior~Member,~IEEE,}
        Umberto~Spagnolini,~\IEEEmembership{Senior~Member,~IEEE}%
\thanks{The authors are with the  Department of Electronics, Information and Bioengineering (DEIB) of Politecnico di Milano, 20133 Milan, Italy  (e-mail: [name.surname]@polimi.it}
}

\begin{document}

\maketitle

\begin{abstract}
This paper presents an echo-side modulation framework for integrated sensing and communication (ISAC) systems. A space-time reconfigurable intelligent surface (ST-RIS) impresses a continuous-phase modulation onto the radar echo, enabling uplink data transmission with a phase modulation of the transmitted radar-like waveform. The received signal is a multiplicative composition of the sensing waveform and the phase for communication. Both functionalities share the same physical signal and perceive each other as impairments. 

The achievable communication rate is expressed as a function of a coupling parameter that links sensing accuracy to phase error accumulation.
Under a fixed bandwidth constraint, the sensing and communication figures of merit define a convex Pareto frontier. The optimal bandwidth allocation satisfying a minimum sensing requirement is derived in closed form. The modified Cramér-Rao bound (MCRB) for range estimation is derived in closed form; this parameter must be estimated to compensate for the frequency offset before data demodulation. Frame synchronization is formulated as a generalized likelihood ratio test (GLRT), and the detection probability is obtained through characteristic function inversion, accounting for residual frequency errors from imperfect range estimation. Numerical results validate the theoretical bounds and characterize the trade-off across the operating range.
\end{abstract}

\section{Introduction}
\label{sec:introduction}

Integrated sensing and communication (ISAC) has emerged as a foundational design paradigm for sixth-generation wireless systems. By sharing spectrum, hardware, and often the transmitted waveform between radar sensing and data communication, ISAC addresses spectrum scarcity, reduces infrastructure cost, and enables applications where both functionalities are inherently coupled~\cite{Liu_JSAC2022, Zhang_JSTSP2021, Waldschmidt_PIEEE2021, Tagliaferri_Access2023, mura2025optimized}.

ISAC system design admits two complementary philosophies. Communication-centric approaches use a communication waveform and extract sensing information as a secondary functionality. Sensing-centric approaches optimize the waveform for radar performance and embed communication in ways that minimally perturb the sensing function~\cite{Liu_Survey2022}. A further distinction concerns where communication embedding occurs. The dominant approach embeds data at the transmitter, modifying the radar waveform to carry information symbols~\cite{liu2018toward, Zhang2022metaradar}. This transmit-side embedding couples the two functionalities at the source, requiring careful waveform co-design to manage mutual interference. An alternative, less explored approach embeds communication on the echo side: the transmitted waveform remains a pure sensing signal, and information is impressed upon the reflected echo by a device in the environment. This echo-side embedding preserves the integrity of the sensing waveform while enabling uplink communication from the reflecting device back to the radar receiver. The framework developed herein adopts this philosophy.

Programmable metasurfaces provide an enabling technology for echo-side ISAC. A metasurface consists of a planar array of sub-wavelength unit cells whose electromagnetic response can be individually configured~\cite{basar2019wireless, DiRenzo_JSAC2020, Cui_LSA2014}. When the surface impedance varies spatially, the tangential momentum of incident waves is not conserved, enabling beam steering toward arbitrary directions~\cite{Yu_Science2011}. When the impedance is modulated in time, temporal symmetry is broken, and the incident wave undergoes frequency conversion~\cite{Hadad_PRB2015, Ramaccia_TAP2020}.

Two paradigms govern temporal modulation. Space-time coding (STC) applies periodic switching sequences, generating harmonics at integer multiples of the modulation frequency~\cite{Zhang_NatComm2018, Dai_TAP2020}. Space-time modulation (STM) employs general time-varying control signals, enabling continuous phase evolution~\cite{Mizmizi_JSAC2024, Tabrizi_NatElectron2021}. This distinction is consequential: STC suits harmonic manipulation and spectral spreading, while STM supports coherent modulation schemes requiring continuous phase control. The framework developed herein accommodates general continuous-phase modulation (CPM) formats; we focus on CPM as a representative class that enables coherent detection while maintaining spectral efficiency.

Prior work on metasurface-enabled ISAC falls into three categories, none addressing echo-side embedding with coherent modulation. The dominant approach treats the metasurface as a passive element shaping propagation to benefit both functionalities. Liu~\emph{et~al.}~\cite{Liu_JSTSP2022} jointly optimize dual-functional waveform and RIS reflection to maximize radar SINR under communication constraints. Wang~\emph{et~al.}~\cite{Wang_TVT2021} extend this to discrete-phase RIS with Cram\'{e}r-Rao bound (CRB) constraints. Subsequent work incorporates active beamforming and multi-metric optimization~\cite{Xu_TCOM2024}. In these systems, the RIS assists both functionalities but carries no data.

The second category exploits harmonic generation in STC metasurfaces for concurrent dual-function operation. Chen~\emph{et~al.}~\cite{Chen_NatComm2025} demonstrate schemes where different frequency components serve communication and sensing, achieving direction-of-arrival estimation alongside data transmission. This forward-link approach embeds data in the metasurface's radiated field rather than in reflected echoes.

A third category explores echo-side embedding through backscatter. Symbiotic radio~\cite{Liang_TCCN2020} enables passive devices to modulate incident signals for spectrum-efficient communication without active RF chains. This paradigm extends to ISAC: Tu~\emph{et~al.}~\cite{Tu_TWC2024} and Tao~\emph{et~al.}~\cite{Tao_TWC2024} analyze RIS-enabled backscatter with joint sensing and communication. In parallel, radar-centric systems using Van~Atta arrays achieve centimeter-level localization with FMCW radars~\cite{Lu_SIGCOMM2023}, and BiScatter~\cite{Okubo_SIGCOMM2024} demonstrates two-way communication through chirp-slope modulation. However, these systems employ simple modulation (OOK, BPSK), and theoretical frameworks coupling sensing accuracy to communication throughput for coherent higher-order modulation remain undeveloped.

This paper considers an ST-RIS that recycles radar illumination to realize coherent uplink communication embedded in the sensing echo. The ST-RIS applies time-varying phase modulation to its reflection coefficient, synthesizing an apparent Doppler shift superimposed on any physical Doppler. By encoding symbols in this apparent Doppler, the metasurface transmits data without generating a carrier. The radar receiver observes an echo from the ST-RIS containing two information components: the beat frequency, which encodes the ST-RIS range, and the apparent Doppler impressed by the time-varying reflection coefficient, which carries the communication data. In this framework, the ST-RIS serves a dual role: it is simultaneously a cooperative target whose range must be estimated for signal demodulation, and a communication endpoint transmitting data back to the radar.

This architecture offers several advantages. The sensing waveform remains unmodified, preserving range resolution and ambiguity function. The ST-RIS consumes minimal power by modulating an existing signal rather than generating one, and coherent communication with higher spectral efficiency than the binary keying used in conventional backscatter.

Realizing this vision presents fundamental challenges. Sensing and communication share the same physical signal, each perceiving the other as an impairment. From a communication perspective, the ST-RIS range introduces frequency and phase offsets that require estimation before symbol detection. From the sensing perspective, communication modulation appears as phase noise, which can potentially degrade range accuracy. These challenges motivate a rigorous theoretical treatment.

This paper develops a theoretical framework for sensing-centric ISAC systems employing space-time modulated metasurfaces for coherent backscatter communication. The main contributions are:

\begin{enumerate}
    \item We formulate a novel framework for communication superposed to radar echoes via ST-RIS phase modulation. The framework decomposes the metasurface phase into spatial and temporal components, which induce a multiplicative coupling between sensing and communication, and accommodates clutter interference.
    
    \item We derive the performance on range estimation, the parameter governing frequency compensation at the receiver, obtaining a closed-form expression that explicitly quantifies how modulation parameters influence sensing accuracy. 
    
    \item We formulate frame synchronization as a generalized likelihood ratio test (GLRT) and derive the detection and false alarm probabilities through characteristic function analysis, accounting for residual frequency offset from imperfect range estimation.
    
    \item We analyze the achievable rate under phase error accumulation caused by range estimation uncertainty, deriving a closed-form expression for the degradation parameter that couples sensing accuracy to communication throughput.
    
    \item We characterize the Pareto frontier between sensing and communication performance, parametrized by the symbol period, and derive the optimal operating point under a sensing constraint.
\end{enumerate}

The remainder of this paper is organized as follows. Section~\ref{sec:system} presents the system model. Section~\ref{sec:modulation} develops the echo-side modulation framework. Section~\ref{sec:performance} derives fundamental performance bounds for sensing and communication. Section~\ref{sec:optimization} addresses the sensing-communication trade-off. Section~\ref{sec:results} presents numerical results, and Section~\ref{sec:conclusions} concludes the paper.
\section{System Model}\label{sec:system}

\begin{figure}[b]
    \centering
    \includegraphics[width=0.8\linewidth]{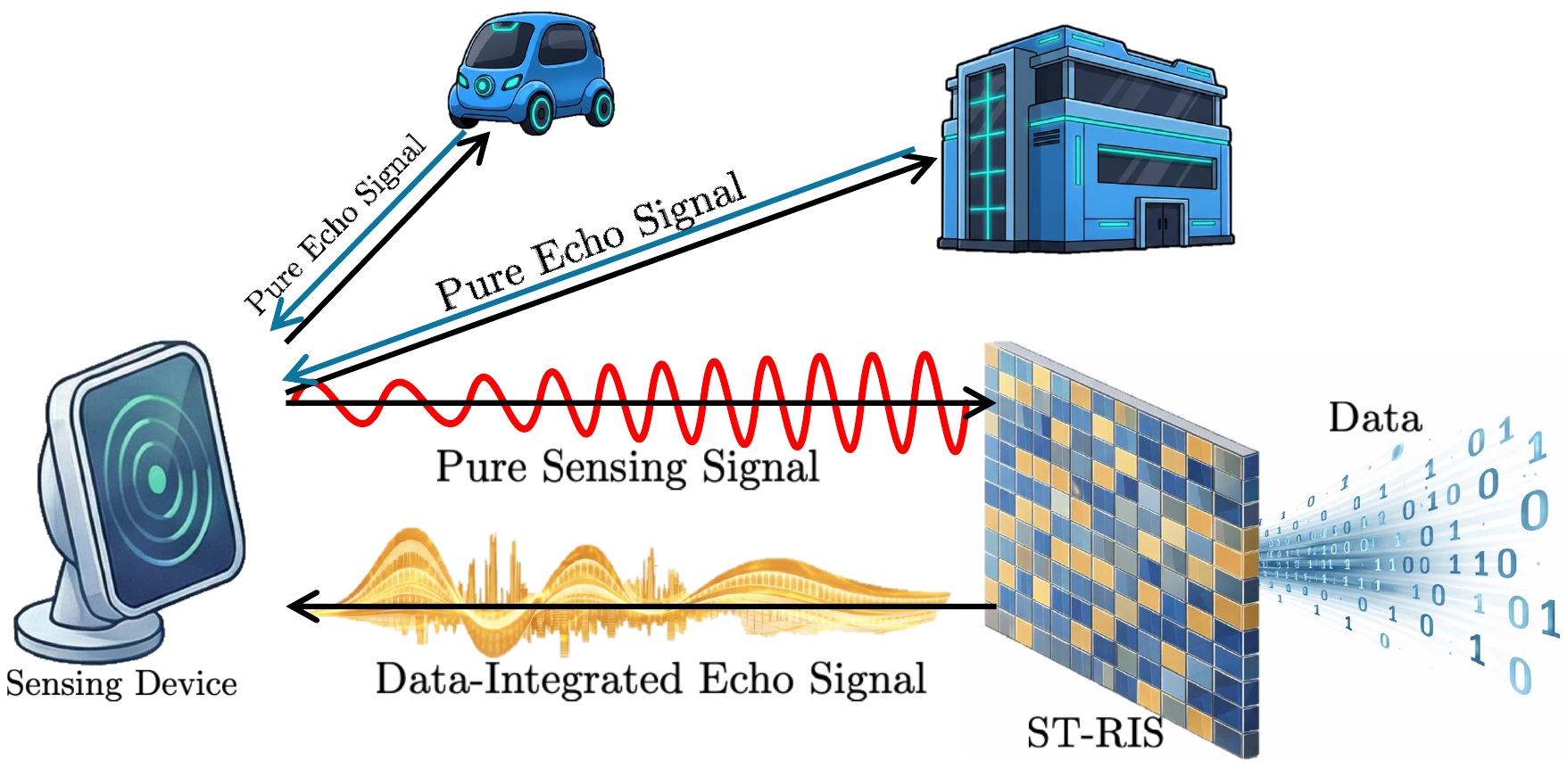}
    \caption{Echo-side ISAC scenario. A Sensing device illuminates an ST-RIS and $Q$ clutter targets. The ST-RIS communicates by impressing a time-varying phase modulation to the reflected signal.}
    \label{fig:scenario}
\end{figure}

Consider the scenario depicted in Fig.~\ref{fig:scenario}, where a radar is equipped with $N_t$ transmit and $N_r$ receive antenna elements, with phase center located at $\mathbf{p}_{\rm r} = [x_{\rm r}, y_{\rm r}, z_{\rm r}]^{\sf T}$. Within the environment, an ST-RIS is deployed at position $\mathbf{p}_{\rm s} = [x_{\rm s}, y_{\rm s}, z_{\rm s}]^{\sf T}$, at distance $R_s = \|\mathbf{p}_{\rm r} - \mathbf{p}_{\rm s}\|_2$ from the radar. The ST-RIS comprises $M = M_x \times M_y$ meta-atoms arranged in a uniform planar array (UPA) with inter-element spacing $d = \lambda_c/4$, where $\lambda_c = c/f_c$ denotes the carrier wavelength.
In addition to the ST-RIS, $Q$ scattering objects are present in the environment. The $q$-th scatterer is located at $\mathbf{p}_q = [x_q, y_q, z_q]^{\sf T}$, at distance $R_q = \|\mathbf{p}_{\rm r} - \mathbf{p}_q\|_2$ from the radar, and moves with radial velocity $v_q$. 

Throughout this work, we assume the ST-RIS to be static. Any physical motion of the ST-RIS would manifest as a geometric Doppler shift superimposed on the communication-induced apparent Doppler, and can be incorporated straightforwardly into the signal model.

\subsection{Transmitted Signal}\label{subsec:tx_signal}

The radar transmits a coherent pulse train. The continuous-time signal radiated by the $n$-th transmit antenna is
\begin{equation}\label{eq:tx_signal}
    s_n(t) = \sqrt{\frac{P_t}{N_t}} \sum_{i=-\infty}^{\infty} p_n(t - iT_{\rm pri}) \, e^{j2\pi f_c t},
\end{equation}
where $P_t$ denotes the total transmitted power, $p_n(t)$ is the baseband waveform with bandwidth $B$, $T_{\rm pri}$ is the pulse repetition interval (PRI), and $f_c$ is the carrier frequency. The waveform $p_n(t)$ is normalized such that $\int |p_n(t)|^2 \, dt = 1$.

\subsection{ST-RIS Reflection Model}\label{subsec:ris_model}

The ST-RIS modulates the incident electromagnetic field through a space-time-varying reflection coefficient matrix. Upon detecting the incident radar pulse, the ST-RIS performs spatial alignment to configure the reflection toward the radar, then initiates phase modulation after a time offset $\tau_c$ relative to pulse arrival. This offset accounts for acquisition and processing latency at the ST-RIS and is unknown to the Rx.

Let $\beta_{m_x, m_y}(t)$ denote the information bearing time-varying phase applied to the meta-atom at position $(m_x, m_y)$ in the array grid, with $m_x \in \{0, \ldots, M_x-1\}$ and $m_y \in \{0, \ldots, M_y-1\}$. The time-varying reflection coefficient matrix
\begin{equation}\label{eq:ref_matrix}
    \boldsymbol{\Phi}(t) = \mathrm{diag}\left( \boldsymbol{\phi}(t) \right) \in \mathbb{C}^{M \times M},
\end{equation}
is assumed diagonal as in \cite{basar2019wireless}. The reflection coefficients are
\begin{equation}\label{eq:ref_vector}
    \boldsymbol{\phi}(t) = \left[ \phi_{0,0}(t), \ldots, \phi_{M_x-1, M_y-1}(t) \right]^{\sf T},
\end{equation}
where each element can be expressed as
\begin{equation}\label{eq:ref_element}
    \phi_{m_x, m_y}(t) = \frac{\alpha(\boldsymbol{\theta}_{\rm F}, \boldsymbol{\theta}_{\rm B})}{M} \, e^{j\beta_{m_x, m_y}(t)}.
\end{equation}
Here, $\alpha(\boldsymbol{\theta}_{\rm F}, \boldsymbol{\theta}_{\rm B})$ captures the angular dependence of the meta-atom response as a function of the forward angles $\boldsymbol{\theta}_{\rm F}$ and backward angles $\boldsymbol{\theta}_{\rm B}$, and can be modeled as in \cite{bjornson2020}. 
The normalization factor $1/M$ in~\eqref{eq:ref_element} ensures that $\|\boldsymbol{\Phi}(t)\|_{\rm F}^2 = 1$, decoupling the per-element response from the array size. When the meta-atom phases are aligned for coherent reflection, the aggregate reflected field scales linearly with $M$; the resulting $M^2$ power gain at the receiver arises from the coherent summation of $M$ unit-amplitude contributions, and it is consistent with standard RIS models~\cite{basar2019wireless}.

The phase function $\beta_{m_x, m_y}(t)$ embeds both a spatial component, which steers the reflected beam, and a temporal component, which modulates the data onto the echo. If the angles of arrival $\boldsymbol{\theta}_{\rm F} =(\vartheta_{\rm F}, \phi{\rm F})$ are known, the ST-RIS can be configured for retroreflection, directing the reflected signal back toward the radar. The acquisition of these angles can be performed through standard beam training procedures based on codebook sweeping~\cite{Wang_TCOM2023_BeamTraining}; the specific structure of $\beta_{m_x, m_y}(t)$ and its role in communication are detailed in Section~\ref{sec:modulation}.

\subsection{Channel Model}\label{subsec:channel_model}

At millimeter-wave and sub-THz frequencies, the propagation environment exhibits high path loss and sparse scattering. For the backscatter link under consideration, the signal traversing the radar$\rightarrow$ST-RIS$\rightarrow$radar path suffers attenuation proportional to $R_s^4$, while paths involving additional reflections experience attenuation of order $R^6$ or higher~\cite{li2008mimo}. Accordingly, we adopt a line-of-sight (LoS) channel model.

Let $\mathbf{h}_n^{\rm F}(t) \in \mathbb{C}^{M \times 1}$ denote the forward channel impulse response from the $n$-th transmit antenna to the ST-RIS, and let $\mathbf{h}_m^{\rm B}(t) \in \mathbb{C}^{1 \times M}$ denote the backward channel impulse response from the ST-RIS to the $m$-th receive antenna. Under LoS propagation, the $(i,j)$-th element of these channel vectors takes the form
\begin{equation}\label{eq:channel_element}
    h_{i,j}(t) = \frac{1}{\sqrt{N_{\rm a} M}} \, e^{j\psi} \, \delta\left( t - \tau_{i,j} \right),
\end{equation}
where $N_{\rm a} \in \{N_t, N_r\}$ denotes the number of radar antennas involved in the link. The term $e^{j\psi}$ represents a constant phase offset arising from residual hardware impairments after calibration; $\psi$ is modeled as uniformly distributed over $[0, 2\pi)$ across realizations. The normalization by $\sqrt{N_{\rm a} M}$ ensures that $\|\mathbf{H}\|_{\rm F}^2 = 1$ for any channel matrix $\mathbf{H}$.

The propagation delay $\tau_{i,j}$ captures both the delay between phase centers $\tau$ and the excess delays introduced by the antenna $\Delta\tau_i^{\rm ant}$ and meta-atom positions $\Delta\tau_j^{\rm ris}$:
\begin{equation}\label{eq:delay_decomposition}
    \tau_{i,j} = \tau + \Delta\tau_i^{\rm ant} + \Delta\tau_j^{\rm ris},
\end{equation}
where $\tau = R_s/c$ is the delay between the radar and ST-RIS phase centers, and $\Delta\tau_i^{\rm ant}$, $\Delta\tau_j^{\rm ris}$ account for the spatial offsets of the $i$-th antenna element and $j$-th meta-atom, respectively. 

For the clutter targets, the channel between the $n$-th transmit antenna and $m$-th receive antenna via the $q$-th scatterer is
\begin{equation}\label{eq:clutter_channel}
    h_{m,n}^{(q)}(t) = \frac{1}{\sqrt{N_t N_r}} \, e^{j\psi_{m,n}^{(q)}} \, \delta\left( t -2\, \tau_q(t) \right),
\end{equation}
where $\tau_q(t) = R_q(t)/c$ is the $q$-th propagation delay.

\subsection{Received Signal}\label{subsec:rx_signal}

\begin{figure}[b]
    \centering
    \includegraphics[width=0.7\columnwidth]{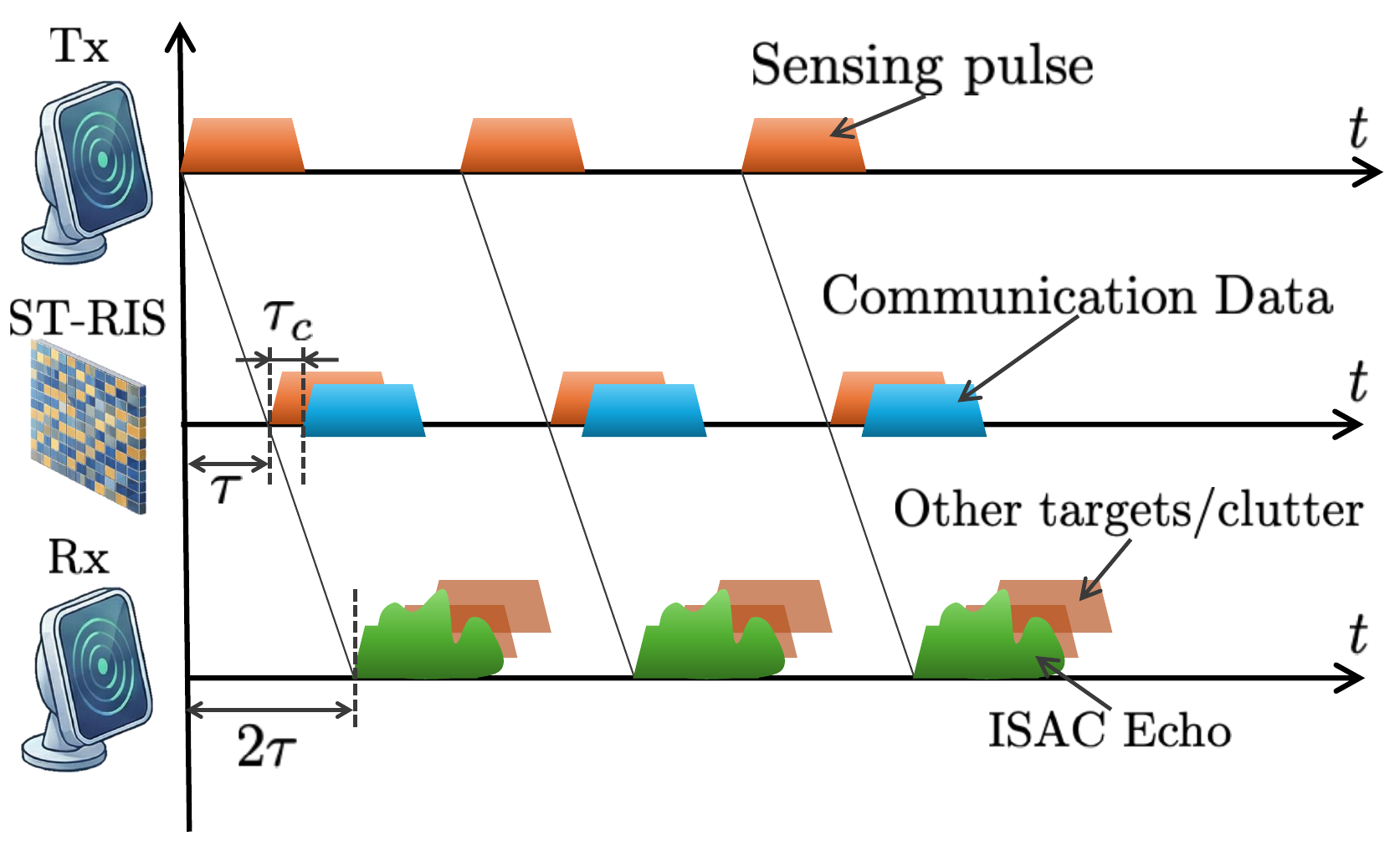}
    \caption{Timing diagram of echo-side ISAC. The radar transmits a sensing pulse (orange) that reaches the ST-RIS after one-way delay $\tau$. The ST-RIS applies phase modulation encoding data (blue), starting at offset $\tau_c$ relative to pulse arrival. At the receiver, the modulated echo (green) arrives after round-trip delay $2\tau$ and combines with unmodulated clutter returns (light orange).}
    \label{fig:timing_diagram}
\end{figure}

In echo-side ISAC, the transmitter radiates a conventional sensing waveform; the dual-function signal arises only at the receiver, where the ST-RIS-modulated echo carries both range information and communication data (Fig.~\ref{fig:timing_diagram}). Demodulating the data requires estimating two parameters: the propagation delay $\tau$, which determines the frequency offset induced by the chirp waveform, and the modulation offset $\tau_c$, which governs symbol boundaries. This task is complicated by the presence of unmodulated reflections from environmental scatterers, which appear as interference.

To capture this structure, we decompose the received signal into three contributions:
\begin{equation}\label{eq:rx_signal}
    y_m(t) = y_m^{\rm ris}(t) + y_m^{\rm clut}(t) + z_m(t).
\end{equation}
The first term represents the ST-RIS echo:
\begin{equation}\label{eq:rx_ris}
    y_m^{\rm ris}(t) = \sqrt{\varrho_s} \sum_{n=1}^{N_t} \mathbf{h}_m^{\rm B}(t) * \boldsymbol{\Phi}(t) \left( \mathbf{h}_n^{\rm F}(t) * s_n(t) \right),
\end{equation}
where $*$ denotes convolution and $\varrho_s$ is the received power from the ST-RIS path. Following the radar range equation for RIS-assisted links~\cite{bjornson2020},
\begin{equation}\label{eq:power_ris}
    \varrho_s = \frac{P_t G_t G_r \lambda_c^2 M^2}{(4\pi)^3 R_s^4},
\end{equation}
where the factor $M^2$ reflects the coherent aperture gain. The time-varying reflection matrix $\boldsymbol{\Phi}(t)$ introduces a modulation onto the reflected signal; as developed in Section~\ref{sec:modulation}, this modulation manifests as an apparent Doppler shift carrying communication information. The second term captures the aggregate clutter contribution:
\begin{equation}\label{eq:rx_clutter}
    y_m^{\rm clut}(t) = \sum_{n=1}^{N_t} \sum_{q=1}^{Q} \sqrt{\varrho_q} \, h_{m,n}^{(q)}(t) * s_n(t),
\end{equation}
where $\varrho_q = P_t G_t G_r \lambda_c^2 \sigma_q / [(4\pi)^3 R_q^4]$ is the received power from the $q$-th scatterer with RCS $\sigma_q$. Finally, $z_m(t) \sim \mathcal{CN}(0, \sigma_z^2)$ is thermal noise with power $\sigma_z^2$.

\section{Echo-Side Modulation Design}\label{sec:modulation}

The objective of the proposed framework is to employ the ST-RIS as a nearly passive device capable of delivering communication data to the radar terminal by modulating the reflected echo. In this section, we develop the signal model for the ST-RIS-modulated return and examine its implications for both sensing and communication functions.

\subsection{Simplified Radar Signal Model}\label{subsec:simplified_model}

To illustrate the essential structure of the received signal, we first consider a simplified scenario with a single transmit antenna ($N_t = 1$), a single receive antenna ($N_r = 1$), and no clutter ($Q = 0$). These assumptions will be relaxed in Section~\ref{subsec:general_model}. Under these conditions, the received signal in~\eqref{eq:rx_signal} reduces to
\begin{equation}\label{eq:rx_signal_simple}
    y(t) = \sqrt{\varrho_s} \, \mathbf{h}^{\rm B}(t) * \boldsymbol{\Phi}(t) \left( \mathbf{h}^{\rm F}(t) * s(t) \right) + z(t).
\end{equation}
The phase applied at each meta-atom is decomposed into a spatial $\varphi_{m_x, m_y}$ and a temporal $\gamma(t)$ component, which for the $(m_x, m_y)$-th meta-atom is expressed as
\begin{equation}\label{eq:st_phase}
    \beta_{m_x, m_y}(t) = \varphi_{m_x, m_y} + \gamma(t),
\end{equation}
for $m_x = 0, \ldots, M_x - 1$ and $m_y = 0, \ldots, M_y - 1$.
Expanding~\eqref{eq:rx_signal_simple} with the phase decomposition~\eqref{eq:st_phase} and evaluating the convolutions yields (\ref{eq:rx_signal_expanded}, in the next page),
\begin{figure*}[!t]
\begin{equation}\label{eq:rx_signal_expanded}
\begin{split}
    y(t) &= \frac{\mu_s}{M} \sum_{m_x=0}^{M_x-1} \sum_{m_y=0}^{M_y-1} e^{j 4\pi f_c \Delta t_{m_x, m_y}} \, e^{j \varphi_{m_x, m_y}} \, e^{j \gamma(t - \Delta t_{m_x, m_y} - \tau)} \, s(t - 2\Delta t_{m_x, m_y} - 2\tau) + z(t) \\[4pt]
    &\overset{(a)}{\approx} \frac{\mu_s}{M} \, s(t - 2\tau) \sum_{m_x=0}^{M_x-1} \sum_{m_y=0}^{M_y-1} e^{j 4\pi f_c \Delta t_{m_x, m_y}} \, e^{j \varphi_{m_x, m_y}} \, e^{j \gamma(t - \Delta t_{m_x, m_y} - \tau)} + z(t) \\[4pt]
    &\overset{(b)}{\approx} \frac{\mu_s}{M} \, s(t - 2\tau) \, e^{j \gamma(t - \tau)} \sum_{m_x=0}^{M_x-1} \sum_{m_y=0}^{M_y-1} e^{j 4\pi f_c \Delta t_{m_x, m_y}} \, e^{j \varphi_{m_x, m_y}} + z(t),
\end{split}
\end{equation}
\hrulefill
\end{figure*}
where $\mu_s = e^{j 2\psi} \sqrt{\varrho_s}$ is a complex coefficient that aggregates the round-trip channel phase $2\psi$ and the path loss. %
The term
\begin{equation}\label{eq:delay_metaatom}
    \Delta t_{m_x, m_y} = \frac{1}{c} \left( \frac{\mathbf{p}_{m_x, m_y}^{\sf T} \mathbf{k}}{\|\mathbf{k}\|} \right) = m_x \Delta t_x + m_y \Delta t_y
\end{equation}
represents the excess one-way propagation delay associated with the $(m_x, m_y)$-th meta-atom, where $\mathbf{p}_{m_x, m_y} = [m_x, m_y, 0]^{\sf T}\; d$ is the position of the meta-atom in local ST-RIS coordinates, $\mathbf{k}$ is the forward wavevector, and
\begin{equation}\label{eq:delay_components}
    \Delta t_x = \frac{d \cos\vartheta \cos\phi}{c}, \qquad \Delta t_y = \frac{d \cos\vartheta \sin\phi}{c}
\end{equation}
are the inter-element delays along the $x$ and $y$ directions, respectively. The angles $\vartheta$ and $\phi$ denote the elevation and azimuth of the forward wave.

Let $\Delta T$ denote the maximum differential delay across the ST-RIS aperture:
\begin{equation}\label{eq:max_delay}
    \Delta T = (M_x - 1) \Delta t_x + (M_y - 1) \Delta t_y.
\end{equation}
The approximations in~\eqref{eq:rx_signal_expanded} rely on the following conditions:

\emph{Approximation~(a):} The radar waveform $s(t)$ does not exhibit spatially wideband effects across the ST-RIS aperture, i.e., $s(t - 2\Delta T) \approx s(t)$. This condition is satisfied when
\begin{equation}\label{eq:approx_a}
    2 \Delta T \ll \frac{1}{B},
\end{equation}
where $B$ is the bandwidth of the transmitted waveform. Under this assumption, beam squinting effects are negligible~\cite{Wang2018_widebandBF}.

\emph{Approximation~(b):} The information-bearing phase signal $\gamma(t)$ remains approximately constant over the aperture delay, i.e., $\gamma(t - \Delta T) \approx \gamma(t)$. This requires
\begin{equation}\label{eq:approx_b}
    \Delta T \ll T_c,
\end{equation}
where $T_c$ is the symbol period of the communication signal.

Under approximations~(a) and~(b), the spatial and temporal components of the phase can be treated independently. The spatial phase $\varphi_{m_x, m_y}$ is then designed to maximize the coherent reflection gain. Specifically, setting
\begin{equation}\label{eq:spatial_phase}
    \varphi_{m_x, m_y} = -\frac{4\pi d}{\lambda_c} \left( m_x \cos\vartheta \cos\phi + m_y \cos\vartheta \sin\phi \right)
\end{equation}
compensates for the geometric phase term $e^{j 4\pi f_c \Delta t_{m_x, m_y}}$ in~\eqref{eq:rx_signal_expanded}, thereby aligning the contributions from all meta-atoms in phase, which yields
\begin{equation}\label{eq:rx_multiplicative}
    y(t) \approx \mu_s \, e^{j \gamma(t - \tau)} \, s(t - 2\tau) + z(t).
\end{equation}
The phase compensation in~\eqref{eq:spatial_phase} requires knowledge of the forward and backward angles $\boldsymbol{\theta}_{\rm F/B}$. These angles can be acquired via codebook-based beam training procedures~\cite{Wang_TCOM2023_BeamTraining}; once obtained, the ST-RIS controller configures the spatial phase accordingly.

\subsection{Sensing Operation}\label{subsec:sensing_design}

To illustrate the interaction between sensing and communication, we consider a linear frequency-modulated (LFM) chirp waveform, widely employed in radar systems \cite{li2008mimo}. The baseband waveform is defined as
\begin{equation}\label{eq:lfm_waveform}
    p(t) = e^{j \pi \kappa t^2}, \quad t \in [0, T_{\rm pri}),
\end{equation}
where $\kappa = B / T_{\rm pri}$ denotes the chirp rate.
In conventional radar processing, the received signal is first down-converted to baseband by carrier removal, then mixed with a replica of the transmitted waveform to produce the beat signal. This is called dechirping, when applied to the backscattered communication signal~\eqref{eq:rx_multiplicative} yields
\begin{equation}\label{eq:beat_signal}
    d(t) = p^*(t) \, y(t) = \mu_s \, e^{j \gamma(t - \tau)} \, e^{j (2\pi f_\tau t + \theta_\tau)} + \tilde{z}(t),
\end{equation}
where $\tilde{z}(t) = p^*(t) z(t)$ is the filtered noise, and $f_\tau = 2 \kappa \tau $ is the beat frequency, and $\theta_\tau = 4\pi \kappa \tau^2 - 4\pi f_c \tau$ is a constant phase offset comprising the residual video phase and the carrier phase accumulated over the round-trip path.

The structure of~\eqref{eq:beat_signal} reveals the fundamental coupling between sensing and communication in the proposed framework. From a sensing perspective, the range information is encoded in the beat frequency $f_\tau= 2 \kappa \tau $, which can be extracted via spectral analysis. However, the multiplicative term $e^{j \gamma(t - \tau)}$ introduces a time-varying phase modulation that acts as interference on the sensing signal. From a communication perspective, the information signal $\gamma(t - \tau)$ is corrupted by a frequency offset $f_\tau$ and a phase offset $\theta_\tau$, both of which must be compensated for data recovery.

\subsection{Generalized Scenario $(N_t>1, N_r>1, Q>0)$}\label{subsec:general_model}

We now extend the signal model to multiple antennas and clutter. The beat signal for the $(n,m)$-th transmit-receive pair is
\begin{align}\label{eq:beat_signal_general}
    d_m^n(t) &= \mu_s \, e^{j \gamma(t - \tau)} \, e^{j (2\pi f_\tau t + \theta_\tau + \varphi_{nm})} \\ \notag 
    &+ \sum_{q=1}^{Q} \mu_q \, e^{j 2\pi (f_{\tau_q} + f_{D,q}) t + \varphi_{nm}^{(q)}} + \tilde{z}_m(t),
\end{align}
where $\varphi_{nm} = 2\pi f_c (\Delta\tau_n^{\rm tx} + \Delta\tau_m^{\rm rx})$ captures the spatial phase from antenna positions, enabling angle estimation via array processing. The clutter term represents interference from $Q$ scatterers, with $\mu_q$, $f_{\tau_q}$, and $f_{D,q}$ denoting the complex amplitude, beat frequency, and Doppler shift of the $q$-th target, respectively.

\subsection{Communication Signal Model}\label{subsec:comm_model}

The information-bearing phase signal $\gamma(t)$ is modeled as
\begin{equation}\label{eq:cpm_signal}
    \gamma(t) = 2\pi h \sum_{i=0}^{N_s-1} b_i \, g(t - i T_c - \tau_c),
\end{equation}
where $h$ denotes the modulation index, $\{b_i\}$ is the sequence of transmitted symbols drawn from an alphabet $\mathcal{A}$ with cardinality $|\mathcal{A}| = L$, $T_c$ is the symbol period, $g(t)$ is the phase shaping pulse, and $\tau_c$ is an unknown time offset at which the ST-RIS initiates modulation within the radar PRI. The symbols are independent and uniformly distributed over $\mathcal{A}$.

The choice of $g(t)$ determines the modulation format within the CPM family. For example, setting $g(t) = (t/2T_c)\,\mathrm{rect}(t/T_c)$ produces continuous-phase frequency-shift keying (CP-FSK). More general choices, such as raised-cosine or Gaussian pulses, yield smoother phase trajectories with improved spectral properties~\cite{anderson2013digital}. Throughout the analysis, we consider general CPM formats; the results specialize to specific schemes by appropriate choice of $g(t)$ and $h$.

The $N_s$ transmitted symbols are organized into a frame comprising a known preamble $\mathbf{b}_p = [b_0, \ldots, b_{L_p-1}]^{\sf T}$ of $L_p$ symbols, followed by $L_d = N_s - L_p$ data symbols. The preamble enables frame synchronization and provides a phase reference for coherent demodulation. 
%
%
\section{Fundamental Performance Limits}
\label{sec:performance}

This section derives the fundamental limits governing ST-RIS range estimation required for frequency compensation and frame synchronization. We first establish the MCRB for ST-RIS range estimation under unknown data symbols, then develop a GLRT for frame synchronization acquisition.

\subsection{ST-RIS Range Estimation Under Unknown Modulation}
\label{sec:mcrb}

The beat signal model in \eqref{eq:beat_signal} provides the foundation for ST-RIS range estimation. After dechirping and sampling at rate $f_s = 1/T_s$, the discrete-time observation at sample index $n$ is
\begin{equation}
  d[n] = \mu_s \, e^{j\Phi[n]} + z[n], \quad n = 0, 1, \ldots, N-1,
  \label{eq:beat_discrete}
\end{equation}
where $\mu_s$ denotes the complex amplitude incorporating path loss and ST-RIS gain, $N = T_0/T_s$ is the total number of samples within the observation window of duration $T_0$, and $z[n] \sim \mathcal{CN}(0, \sigma_z^2)$ represents additive noise with variance $\sigma_z^2 = N_0/T_s$. The instantaneous phase at sample $n$ is:
\begin{equation}
  \Phi[n] = 2\pi f_\tau t_n + \theta_\tau + \gamma(t_n - \tau),
  \label{eq:phase_discrete}
\end{equation}
where the linear term $2\pi f_\tau n\, T_s$ arising from the beat frequency $f_\tau=2\kappa \tau$, and the ST-RIS modulation $\gamma[n - \tau]$ evaluated at the delayed time instant.
The phase offset $\theta_\tau$ includes contributions from the carrier phase accumulated over the round-trip path. In principle, $\theta_\tau$ depends on $\tau$. However, exploiting this dependence for range estimation would require carrier-phase synchronization with sub-wavelength accuracy, which is impractical at high frequencies. Accordingly, we treat $\theta_\tau$ as a nuisance parameter that is statistically independent of $\tau$. This assumption is standard in delay estimation problems~\cite{richards2014fundamentals}.

The parameter of interest is the one-way propagation delay $\tau = R_s/c$. However, the data symbol sequence $\mathbf{b} = [b_0, b_1, \ldots, b_{N_s-1}]^T$ embedded in $\gamma(\cdot)$ constitutes a nuisance parameter that must be addressed in the estimation procedure. We adopt the MCRB framework~\cite{d1994modified}, which provides a lower bound on the variance of any unbiased estimator when nuisance parameters are present.

For a deterministic parameter $\tau$ in the presence of random nuisance parameters $\mathbf{b}$ with known prior distribution, the MCRB is defined as
\begin{equation}
  \mathrm{MCRB}(\tau) \triangleq \frac{1}{\mathbb{E}_{\mathbf{b}}\bigl[\mathcal{I}(\tau \,|\, \mathbf{b})\bigr]},
  \label{eq:mcrb_def}
\end{equation}
where $\mathcal{I}(\tau \,|\, \mathbf{b})$ denotes the Fisher Information for $\tau$ conditioned on a specific realization of~$\mathbf{b}$.

Under the AWGN model~\eqref{eq:beat_discrete}, the conditional Fisher Information takes the form
\begin{equation}
  \mathcal{I}(\tau \,|\, \mathbf{b}) 
  = \frac{2|\mu_s|^2}{\sigma_z^2} \sum_{n=0}^{N-1} \left(\frac{\partial \Phi[n]}{\partial \tau}\right)^2.
  \label{eq:fisher_conditional}
\end{equation}
Computing the partial derivative of~\eqref{eq:phase_discrete} with respect to $\tau$ yields
\begin{equation}
  \frac{\partial \Phi[n]}{\partial \tau} 
  = 4\pi\kappa t_n  - \dot{\gamma}(t_n - \tau),
  \label{eq:phase_derivative}
\end{equation}
where $\dot{\gamma}(t) = d\gamma/dt$ denotes the instantaneous frequency deviation induced by the ST-RIS modulation. For the CPM format, this derivative evaluates to
\begin{equation}
  \dot{\gamma}(t) = 2\pi h \sum_{i=0}^{N_s-1} b_i \, \dot{g}(t - iT_c),
  \label{eq:gamma_dot}
\end{equation}
with $\dot{g}(t)$ representing the frequency pulse shape.

Substituting \eqref{eq:phase_derivative} into \eqref{eq:fisher_conditional}, we obtain
\begin{equation}
  \mathcal{I}(\tau \,|\, \mathbf{b}) = \frac{2|\mu_s|^2}{\sigma_z^2} \sum_{n=0}^{N-1} \bigl(A_n + B_n\bigr)^2,
  \label{eq:fisher_decomposed}
\end{equation}
where
\begin{align}
  A_n &= 4\pi\kappa t_n, \label{eq:AL_def}\\
  B_n &= -\dot{\gamma}(t_n - \tau) = -2\pi h \sum_{i=0}^{N_s-1} b_i \, \dot{g}(t_n - iT_c - \tau). \label{eq:BL_def}
\end{align}
Expanding the square in~\eqref{eq:fisher_decomposed} produces three terms:
\begin{equation}
  \mathcal{I}(\tau \,|\, \mathbf{b}) = \frac{2|\mu_s|^2}{\sigma_z^2} \Bigl(\mathcal{I}_s + 2\mathcal{I}_{sc}(\mathbf{b}) + \mathcal{I}_c(\mathbf{b})\Bigr),
  \label{eq:fisher_expansion}
\end{equation}
where
\begin{align}
  \mathcal{I}_s &= 16\pi^2\kappa^2 \sum_{n=0}^{N-1} t_n^2, \label{eq:Is_def}\\
  \mathcal{I}_{sc}(\mathbf{b}) &= -8\pi^2\kappa h \sum_{n=0}^{N-1} t_n \sum_{i=0}^{N_s-1} b_i \, \dot{g}(t_n - iT_c - \tau), \label{eq:Isc_def}\\
  \mathcal{I}_c(\mathbf{b}) &= 4\pi^2 h^2 \sum_{n=0}^{N-1} \left(\sum_{i=0}^{N_s-1} b_i \, \dot{g}(t_n - iT_c - \tau)\right)^2. \label{eq:Ic_def}
\end{align}
The term $\mathcal{I}_s$ depends solely on the sensing parameters and observation geometry, while $\mathcal{I}_c(\mathbf{b})$ captures the modulation-induced information and $\mathcal{I}_{sc}(\mathbf{b})$ represents the cross-term.
Taking the expectation over the i.i.d.\ data symbols with $\mathbb{E}[b_i] = 0$ and $\mathbb{E}[b_i^2] = \sigma_b^2$, the cross-term vanishes:
\begin{equation}
  \mathbb{E}_{\mathbf{b}}[\mathcal{I}_{sc}(\mathbf{b})] = 0.
  \label{eq:cross_term_zero}
\end{equation}
For the modulation contribution, under the assumption that inter-symbol interference is negligible (i.e., $\dot{g}(t)$ has support confined to $[0, T_c]$), the double sum simplifies:
\begin{equation}
  \mathbb{E}_{\mathbf{b}}[\mathcal{I}_c(\mathbf{b})] = 4\pi^2 h^2 \sigma_b^2 \sum_{n=0}^{N-1} \sum_{i=0}^{N_s-1} \dot{g}^2(t_n - iT_c - \tau).
  \label{eq:Ic_expectation}
\end{equation}
We now evaluate the sums in closed form. Approximating the discrete sum by an integral and using $N_s = T_0/T_c$ symbols within the observation window:
\begin{equation}
  \mathcal{I}_s = 16\pi^2\kappa^2 \sum_{n=0}^{N-1} t_n^2 
  \approx \frac{16\pi^2\kappa^2}{T_s} \int_0^{T_0} t^2 \, dt 
  = \frac{16\pi^2\kappa^2 T_0^3}{3T_s}.
  \label{eq:Is_closed}
\end{equation}
For CP-FSK with rectangular frequency pulse $\dot{g}(t) = (1/2T_c)\,\mathrm{rect}(t/T_c - 1/2)$, each symbol contributes $T_c/T_s$ samples with $\dot{g}^2 = 1/(4T_c^2)$:
\begin{equation}
  \mathbb{E}_{\mathbf{b}}[\mathcal{I}_c] = 4\pi^2 h^2 \sigma_b^2 \cdot N_s \cdot \frac{T_c}{T_s} \cdot \frac{1}{4T_c^2} 
  = \frac{\pi^2 h^2 \sigma_b^2 N_s}{T_s T_c}.
  \label{eq:Ic_closed}
\end{equation}
Combining~\eqref{eq:Is_closed} and~\eqref{eq:Ic_closed}, the expected Fisher Information becomes
\begin{equation}
  \mathbb{E}_{\mathbf{b}}[\mathcal{I}(\tau \,|\, \mathbf{b})] 
  = \frac{2|\mu_s|^2}{\sigma_z^2} \cdot \frac{\pi^2}{T_s} \left(\frac{16\kappa^2 T_0^3}{3} + \frac{h^2 \sigma_b^2 N_s}{T_c}\right).
  \label{eq:fisher_expected}
\end{equation}
Defining the signal-to-noise ratio (SNR) as $\rho \triangleq |\mu_s|^2 T_0 / N_0$ and substituting $\sigma_z^2 = N_0/T_s$, we obtain, after algebraic manipulation, the MCRB for $\tau$:
\begin{equation}
  \mathrm{MCRB}(\tau) = \frac{3T_0T_c}{2\pi^2 \rho} \cdot \frac{1}{16\kappa^2 T_0^3 T_c + 3h^2\sigma_b^2 N_s}.
  \label{eq:mcrb_final}
\end{equation}

The MCRB~\eqref{eq:mcrb_final} reveals the interplay between sensing and communication parameters. The denominator comprises two terms: $16\kappa^2 T_0^3 T_c$ represents the information contributed by the sensing waveform, while $3h^2\sigma_b^2 N_s$ captures the additional information from the phase modulation. When the observation window contains only the preamble (known symbols with $\sigma_b^2 = 0$), the bound reduces to the standard CRB for ranging.
%
Conversely, when data symbols are present, the modulation contributes positively to range estimation accuracy, provided the receiver can exploit the phase structure. This observation motivates the joint processing approach developed in subsequent sections.

\subsection{Frame Synchronization}\label{subsec:frame_sync}

Reliable demodulation requires accurate knowledge of the frame timing offset $\tau_c$, i.e., when the ST-RIS starts modulating the echo signal. We distinguish this parameter from the propagation delay $\tau$: while $\tau$ determines the signal's time-of-arrival and is estimated for sensing purposes, $\tau_c$ governs symbol boundaries and must be acquired for communication.

The frame structure comprises $L_p$ preamble symbols followed by $L_d$ data symbols. The preamble employs a known sequence $\mathbf{p} = [p_0, p_1, \ldots, p_{L_p-1}]^T$ designed for synchronization, while the data portion carries the information symbols $\mathbf{d} = [d_0, d_1, \ldots, d_{L_d-1}]^T$.

Considering the discrete beat signal in \eqref{eq:beat_discrete}, frame synchronization is formulated as a binary hypothesis test. At each candidate offset $m$, the hypotheses are:
\begin{align}
    \mathcal{H}_0 &: d[n + m] = \mu_s \, e^{j(2\pi f_\tau n T_s + \theta_\tau)} + z[n], \label{eq:H0_tone} \\[3pt]
    \mathcal{H}_1 &: d[n + m] = \mu_s \, e^{j(2\pi f_\tau n T_s + \theta_\tau + \gamma_p[n])} + z[n], \label{eq:H1_preamble}
\end{align}
for $n = 0, \ldots, L_p - 1$. Under $\mathcal{H}_0$, the received signal is a tone at the beat frequency $f_\tau$; under $\mathcal{H}_1$, both the tone at $f_\tau$ and the preamble modulation are present.

Let the tone-matched and preamble-matched templates be
\begin{align}
    s_0[n] &= e^{j 2\pi \hat{f}_\tau n T_s}, \label{eq:tone_template} \\
    s_p[n] &= e^{j(2\pi \hat{f}_\tau n T_s + \gamma_p[n])}, \label{eq:preamble_template}
\end{align}
where $\hat{f}_\tau$ is the beat frequency estimate from the sensing stage. The matched-filter outputs are
\begin{align}
    C_0[m] &= \sum_{n=0}^{L_p - 1} d[n + m] \, s_0^*[n], \label{eq:C0_def} \\
    C_p[m] &= \sum_{n=0}^{L_p - 1} d[n + m] \, s_p^*[n]. \label{eq:Cp_def}
\end{align}
The GLRT statistic, which accounts for unknown amplitude $\mu_s$ and phase $\theta_\tau$, is
\begin{equation}\label{eq:glrt_stat}
    \Lambda[m] = |C_p[m]|^2 - |C_0[m]|^2.
\end{equation}
Under $\mathcal{H}_1$, the preamble correlation dominates, yielding $\Lambda > 0$; under $\mathcal{H}_0$, the tone correlation dominates, yielding $\Lambda < 0$. The joint detection and synchronization decision is
\begin{equation}\label{eq:decision_rule}
    \hat{m} = \arg\max_{m} \Lambda[m], \qquad \Lambda[\hat{m}] \underset{\mathcal{H}_0}{\overset{\mathcal{H}_1}{\gtrless}} \eta,
\end{equation}
where $\eta$ is a detection threshold.

\subsubsection{Effect of Residual Frequency Offset}

Let $\epsilon_f = f_\tau - \hat{f}_\tau$ denote the residual frequency estimation error. Under $\mathcal{H}_1$ at the correct offset $m = m_0$, the preamble-matched output is
\begin{equation}\label{eq:Cp_cfo}
    C_p = \mu_s \, e^{j\psi} \, \chi(\epsilon_f) + \tilde{z}_p,
\end{equation}
where $\psi = \theta_\tau + 2\pi f_\tau m_0 T_s$ is a deterministic phase, and 
\begin{equation}\label{eq:coherence_factor}
    \chi(\epsilon_f) = \sum_{n=0}^{L_p-1} e^{j 2\pi \epsilon_f n T_s} = e^{j\pi \epsilon_f (L_p-1) T_s} \frac{\sin(\pi \epsilon_f L_p T_s)}{\sin(\pi \epsilon_f T_s)}
\end{equation}
is the coherence factor. For $\epsilon_f = 0$, we have $\chi(0) = L_p$. For $\epsilon_f \neq 0$, the magnitude $|\chi(\epsilon_f)| < L_p$, representing a coherent processing loss.
Similarly, the tone-matched output under $\mathcal{H}_1$ becomes
\begin{equation}\label{eq:C0_H1_cfo}
    C_0 = \mu_s \, e^{j\psi} \, \tilde{\Gamma}_p(\epsilon_f) + \tilde{z}_0,
\end{equation}
where
\begin{equation}\label{eq:Gamma_tilde}
    \tilde{\Gamma}_p(\epsilon_f) = \sum_{n=0}^{L_p-1} e^{j(\gamma_p[n] + 2\pi \epsilon_f n T_s)}
\end{equation}
is the preamble DC content modulated by the residual frequency offset.
Under $\mathcal{H}_0$, the roles are exchanged: $C_0 = \mu_s e^{j\psi} \chi(\epsilon_f) + \tilde{z}_0$ and $C_p = \mu_s e^{j\psi} \tilde{\Gamma}_p^*(\epsilon_f) + \tilde{z}_p$.

\subsubsection{Statistical Characterization}

The correlation outputs $C_p$ and $C_0$ are jointly complex Gaussian. However, they are \emph{not} independent: both depend on the same noise samples $z[n]$. The noise components $\tilde{z}_p = \sum_n z[n] s_p^*[n]$ and $\tilde{z}_0 = \sum_n z[n] s_0^*[n]$ have covariance
\begin{equation}\label{eq:noise_cov}
    \mathrm{Cov}(\tilde{z}_p, \tilde{z}_0) = \sigma_z^2 \sum_{n=0}^{L_p-1} e^{-j\gamma_p[n]} = \sigma_z^2 \Gamma_p^*,
\end{equation}
where $\Gamma_p = \sum_n e^{j\gamma_p[n]}$ is the preamble DC content at zero frequency offset.
The correlation coefficient is $\nu_c = \Gamma_p^* / L_p$, with $|\nu_c| \leq 1$. 

The GLRT statistic $\Lambda = |C_p|^2 - |C_0|^2$ is a quadratic form in the jointly Gaussian vector $\mathbf{C} = [C_p, C_0]^{\sf T}$. The exact distribution of $\Lambda$ does not admit a simple closed form; however, the characteristic function can be derived analytically. The detection and false alarm probabilities are then obtained by numerical inversion. The complete derivation is provided in Appendix~\ref{app:glrt_analysis}.

\subsubsection{Detection and False Alarm Probabilities}

Let $\rho = |\mu_s|^2/\sigma_z^2$ denote the per-sample SNR. Assuming negligible correlation ($|\nu_c| \ll 1$), the detection and false alarm probabilities under perfect frequency estimation ($\epsilon_f = 0$) are
\begin{align}
    P_{\rm d} &= \Pr(\Lambda > \eta \mid \mathcal{H}_1) = 1 - F_\Lambda^{(1)}(\eta), \label{eq:Pd_def} \\
    P_{\rm fa} &= \Pr(\Lambda > \eta \mid \mathcal{H}_0) = 1 - F_\Lambda^{(0)}(\eta), \label{eq:Pfa_def}
\end{align}
where $F_\Lambda^{(i)}(\eta)$ denotes the cumulative distribution function (CDF) of $\Lambda$ under $\mathcal{H}_i$. As shown in Appendix~\ref{app:glrt_analysis}, the CDF in \eqref{eq:cdf_formula} is computed via numerical inversion of the characteristic function in~\eqref{eq:cf_lambda}.
For system design purposes, the threshold $\eta$ is determined by the target false alarm probability $P_{\rm fa}^{\rm tgt}$ through numerical inversion of~\eqref{eq:Pfa_def}. The detection probability then follows from~\eqref{eq:Pd_def}.\\
\subsection{Communication Performance}\label{subsec:comL_perf}

After frame synchronization and frequency compensation at Rx, the data symbols from ST-RIS are demodulated. This subsection aims to derive the achievable communication rate and establishes the fundamental trade-off between sensing accuracy and communication throughput.

After compensating the beat frequency $\hat{f}_\tau$ and phase $\hat{\theta}_\tau$, the received signal corresponding to the $k$-th data symbol is
\begin{equation}\label{eq:data_signal}
    r_k = \mu_s \, e^{j(\gamma_d[k] + \phi_k)} + z_k, \quad k = 0, \ldots, L_d - 1,
\end{equation}
where $\gamma_d[k]$ is the data-bearing phase at symbol $k$, and
\begin{equation}\label{eq:phase_error}
    \phi_k = 2\pi \epsilon_f k T_c + \epsilon_\theta
\end{equation}
is the residual phase error, comprising a linearly accumulating term due to the frequency estimation error $\epsilon_f = f_\tau - \hat{f}_\tau$ and a constant phase offset $\epsilon_\theta$.
The frequency error $\epsilon_f$ is a random variable whose variance is bounded by the sensing performance derived in Section~\ref{subsec:sensing_design}:
\begin{equation}\label{eq:freq_error_variance}
    \sigma_\epsilon^2 \triangleq \mathrm{Var}(\epsilon_f) = 4\kappa^2 \, \mathrm{Var}(\hat{\tau}) \geq 4\kappa^2 \cdot \mathrm{MCRB}(\tau).
\end{equation}
The accumulating phase error degrades demodulation performance. To quantify this effect, we treat the phase perturbation as an equivalent interference. For small phase errors ($|\phi_k| \ll 1$), the signal~\eqref{eq:data_signal} can be approximated as
\begin{equation}\label{eq:linearized}
    r_k \approx \mu_s e^{j\gamma_d[k]} + j \mu_s e^{j\gamma_d[k]} \phi_k + z_k.
\end{equation}
The second term acts as multiplicative interference with power $|\mu_s|^2 \mathbb{E}[\phi_k^2]$.
The mean-squared phase error, over $L_d$ data symbols, is
\begin{equation}\label{eq:mean_phase_var}
    \bar{\sigma}_\phi^2 = \frac{1}{L_d} \sum_{k=0}^{L_d-1} \mathbb{E}[\phi_k^2] = (2\pi T_c)^2 \sigma_\epsilon^2 \cdot \frac{(L_d-1)(2L_d-1)}{6} + \sigma_{\epsilon_\theta}^2,
\end{equation}
where $\sigma_{\epsilon_\theta}^2 = \mathrm{Var}(\epsilon_\theta)$. For large $L_d$ and assuming the phase offset is absorbed into the demodulator (or $\sigma_{\epsilon_\theta}^2\ll 1$), the dominant contribution is
\begin{equation}\label{eq:phase_var_approx}
    \bar{\sigma}_\phi^2 \approx \frac{(2\pi T_c \sigma_\epsilon L_d)^2}{3}.
\end{equation}
The effective signal-to-interference-plus-noise ratio (SINR) is
\begin{equation}\label{eq:sinr_eff}
    \rho_{\rm eff} = \frac{|\mu_s|^2}{\sigma_z^2 + |\mu_s|^2 \bar{\sigma}_\phi^2} = \frac{\rho}{1 + \rho \bar{\sigma}_\phi^2},
\end{equation}
where $\rho = |\mu_s|^2 / \sigma_z^2$ is the per-symbol SNR in the absence of phase error.
%
%
Substituting~\eqref{eq:freq_error_variance} and~\eqref{eq:phase_var_approx} into~\eqref{eq:sinr_eff}, we define the \emph{coupling parameter}
\begin{equation}\label{eq:xi_def}
    \xi = \rho \bar{\sigma}_\phi^2 = \frac{16\pi^2 \kappa^2 T_c^2 L_d^2 \rho}{3} \cdot \mathrm{MCRB}(\tau),
\end{equation}
which quantifies the SNR loss due to phase error accumulation. The effective SINR, thus, becomes
\begin{equation}\label{eq:sinr_xi}
    \rho_{\rm eff} = \frac{\rho}{1 + \xi}.
\end{equation}
The parameter $\xi$ captures the coupling between sensing and communication: poor range estimation (large MCRB) leads to large $\xi$ and degraded communication performance. 

For an AWGN channel with effective SINR $\rho_{\rm eff}$, the spectral efficiency is upper-bounded by the Shannon capacity:
\begin{equation}\label{eq:spectral_eff}
    \eta = \log_2(1 + \rho_{\rm eff}) = \log_2\left( 1 + \frac{\rho}{1 + \xi} \right) \quad \text{[bits/symbol]}.
\end{equation}
For a practical modulation with alphabet size $L$, the achievable rate per symbol is at most $\log_2 L$ bits. The effective spectral efficiency is thus
\begin{equation}\label{eq:eff_spectral}
    \eta_{\rm eff} = \min\left\{ \log_2 L, \, \log_2\left( 1 + \frac{\rho}{1 + \xi} \right) \right\}.
\end{equation}
The communication data rate, accounting for the preamble overhead, is
\begin{equation}\label{eq:data_rate}
    R = \frac{L_d}{T_{\rm pri}} \, \eta_{\rm eff} \quad \text{[bits/s]}.
\end{equation}
%
\section{Sensing-Communication Trade-off Optimization}
\label{sec:optimization}

This section addresses the joint optimization of sensing and communication performance under a shared bandwidth constraint, revealing the fundamental trade-off inherent in the proposed echo-side modulation architecture.

\subsection{Bandwidth Constraint and Fisher Information}
\label{sec:tradeoff}

In practical deployments, the total available bandwidth $B_T$ is fixed by regulatory allocation or transceiver hardware limitations. Although the chirp waveform and the CPM modulation are not separated in frequency (they coexist over the same time-frequency support through multiplication in the time domain) the spectral occupancy of the composite signal is governed by the convolution of their individual spectra. Consequently, the total bandwidth satisfies the constraint
\begin{equation}
  B_s + B_c = B_T,
  \label{eq:bw_constraint}
\end{equation}
where $B_s = \kappa \, T_0$ denotes the chirp bandwidth over the observation window and $B_c \approx (1+h)/T_c$ represents the CPM bandwidth according to Carson's rule~\cite{anderson2013digital}. This constraint implies that allocating additional bandwidth to communication necessarily reduces the bandwidth available for sensing.

To evaluate this interplay, we express the Fisher Information contributions from Section~\ref{sec:mcrb} in terms of the bandwidth allocation. The sensing contribution~\eqref{eq:Is_closed}, after incorporating the SNR factor from~\eqref{eq:mcrb_final}, becomes
\begin{equation}
  \mathcal{I}_s = \frac{32\pi^2 \rho}{3} B_s^2 = a \, B_s^2,
  \label{eq:Is_bandwidth}
\end{equation}
where $a = 32\pi^2 \rho / 3$. Similarly, the modulation contribution~\eqref{eq:Ic_closed} can be written as
\begin{equation}
  \mathcal{I}_c = 2\pi^2 \rho h^2 \sigma_b^2 B_c^2 = b \, B_c^2,
  \label{eq:Ic_bandwidth}
\end{equation}
where $b = 2\pi^2 \rho h^2 \sigma_b^2$ and we have used $B_c \approx 1/T_c$ for notational compactness. The ratio of these coefficients is
\begin{equation}
  \frac{a}{b} = \frac{16}{3 h^2 \sigma_b^2}.
  \label{eq:coeff_ratio}
\end{equation}
For typical modulation parameters ($h = 0.5$, $\sigma_b^2 = 1$), the ratio $a/b=21$; for $h = 1$, it reduces to approximately $a/b=5$. In either case, $a > b$: the sensing waveform extracts more Fisher Information per unit bandwidth than the phase modulation. This asymmetry constitutes the fundamental source of the sensing-communication trade-off.
Under the constraint~\eqref{eq:bw_constraint}, the total Fisher Information for delay estimation is
\begin{equation}
  \mathcal{I}_{\mathrm{tot}}(B_c) = a(B_T - B_c)^2 + b \, B_c^2.
  \label{eq:I_total}
\end{equation}
Taking the derivative with respect to $B_c$:
\begin{equation}
  \frac{\partial \mathcal{I}_{\mathrm{tot}}}{\partial B_c} = -2a(B_T - B_c) + 2b \, B_c = 2(a+b)B_c - 2a B_T.
  \label{eq:dI_dBc}
\end{equation}
This derivative is negative for $B_c < a B_T/(a+b)$, a condition satisfied over most of the practical range since $a \gg b$. Consequently, $\mathcal{I}_{\mathrm{tot}}$ decreases monotonically as communication bandwidth increases: maximum sensing performance occurs at $B_c = 0$ (sensing-only mode), while maximum communication rate requires $B_c = B_T$ (communication-only mode).

\subsection{Normalized Performance Metrics}
\label{sec:normalized_fom}

To facilitate optimization and enable dimensionless comparison, we introduce normalized metrics mapping both sensing and communication performance to the interval $[0, 1]$. Let $\beta = B_c / B_T \in [0, 1]$ denote the fraction of bandwidth allocated to communication.
The sensing figure of merit, normalized by the sensing-only baseline, is
\begin{equation}
  \tilde{\mathcal{S}}(\beta) = \frac{\mathcal{I}_{\mathrm{tot}}(\beta)}{\mathcal{I}_{\mathrm{tot}}(0)} = (1-\beta)^2 + \frac{b}{a}\beta^2.
  \label{eq:S_normalized}
\end{equation}
This metric satisfies $\tilde{\mathcal{S}}(0) = 1$ (maximum sensing capability) and $\tilde{\mathcal{S}}(1) = b/a$ (communication-only operation). 
Instead, the communication figure of merit in \eqref{eq:data_rate}, normalized by the maximum achievable rate, is
\begin{equation}
  \tilde{\mathcal{C}}(\beta) = \frac{R(\beta)}{R(1)} = \beta,
  \label{eq:C_normalized}
\end{equation}
where we assume, to first order, that the achievable rate scales linearly with the allocated bandwidth fraction. This approximation holds when the spectral efficiency (bits per Hertz) remains approximately constant across the operating range. The metric satisfies $\tilde{\mathcal{C}}(0) = 0$ (no communication) and $\tilde{\mathcal{C}}(1) = 1$ (maximum communication rate).

The Pareto frontier in the $(\tilde{\mathcal{S}}, \tilde{\mathcal{C}})$ plane characterizes the optimal trade-off between sensing and communication. Eliminating $\beta$ using $\beta = \tilde{\mathcal{C}}$ from~\eqref{eq:C_normalized} and substituting into~\eqref{eq:S_normalized}:
\begin{equation}
  \tilde{\mathcal{S}} = (1 - \tilde{\mathcal{C}})^2 + \frac{b}{a}\tilde{\mathcal{C}}^2 = 1 - 2\tilde{\mathcal{C}} + \left(1 + \frac{b}{a}\right)\tilde{\mathcal{C}}^2.
  \label{eq:pareto_frontier}
\end{equation}
This expression describes a convex parabola opening upward. The frontier has negative slope
\begin{equation}
  \frac{\partial \tilde{\mathcal{S}}}{\partial \tilde{\mathcal{C}}} = -2 + 2\left(1 + \frac{b}{a}\right)\tilde{\mathcal{C}} < 0
  \label{eq:pareto_slope}
\end{equation}
for $\tilde{\mathcal{C}} < a/(a+b)$. Since $a \gg b$ implies $a/(a+b) \approx 1$, the trade-off persists over essentially the entire operating range.

\subsection{Sensing-Constrained Rate Maximization}
\label{sec:sensing_constrained}

For sensing-centric applications such as automotive radar with secondary communication capability, the design objective is to maximize communication throughput while guaranteeing a minimum sensing performance. This requirement leads to the constrained optimization problem:
\begin{equation}
  \begin{aligned}
    \max_{\beta \in [0,1]} \quad & \tilde{\mathcal{C}}(\beta) = \beta \\
    \text{subject to} \quad & \tilde{\mathcal{S}}(\beta) \geq \tilde{\mathcal{S}}_{\min},
  \end{aligned}
  \label{eq:opt_problem}
\end{equation}
where $\tilde{\mathcal{S}}_{\min} \in [b/a, 1]$ specifies the required sensing capability relative to the sensing-only baseline. 
The sensing constraint from~\eqref{eq:S_normalized} can be rewritten as
\begin{equation}
  (1-\beta)^2 + \frac{b}{a}\beta^2 \geq \tilde{\mathcal{S}}_{\min}.
  \label{eq:constraint_expanded}
\end{equation}
Expanding and rearranging yields the quadratic inequality
\begin{equation}
  \left(1 + \frac{b}{a}\right)\beta^2 - 2\beta + (1 - \tilde{\mathcal{S}}_{\min}) \leq 0.
  \label{eq:constraint_quadratic}
\end{equation}
The roots of the corresponding equality are
\begin{equation}
  \beta_{\pm} = \frac{1 \pm \sqrt{1 - (1 + b/a)(1 - \tilde{\mathcal{S}}_{\min})}}{1 + b/a}.
  \label{eq:beta_roots}
\end{equation}
For the constraint to admit a feasible solution, the discriminant must be non-negative, which holds whenever $\tilde{\mathcal{S}}_{\min} \geq b/a$. The feasible region is $\beta \in [0, \beta_+]$, bounded by the larger root.

Since the objective $\tilde{\mathcal{C}}(\beta) = \beta$ is monotonically increasing, the optimal solution lies at the constraint boundary:
\begin{equation}
  \beta^* = \beta_+ = \frac{1 - \sqrt{1 - (1 + b/a)(1 - \tilde{\mathcal{S}}_{\min})}}{1 + b/a}.
  \label{eq:beta_optimal}
\end{equation}
This closed-form expression provides the maximum communication bandwidth allocation satisfying the sensing requirement.

The coefficient ratio $a/b = 16/(3h^2\sigma_b^2)$ governs the severity of the trade-off. Increasing the modulation index $h$ or employing higher-variance symbol constellations (larger $\sigma_b^2$) reduces $a/b$, thereby improving the information efficiency of the modulation relative to the sensing. This asymmetry underscores the inherent cost of embedding communication into a sensing-optimized waveform and motivates the sensing-constrained design philosophy adopted in this work.

Note that, for CPM with a symmetric $L$-ary alphabet $\mathcal{A} = \{-(L-1), -(L-3), \ldots, +(L-1)\}$, the instantaneous phase increment per symbol is $\Delta\phi = \pi h b_i$, yielding a maximum excursion of $|\Delta\phi|_{\max} = \pi h (L-1)$. To ensure unambiguous phase detection and prevent symbol errors due to phase wrapping, the modulation parameters must satisfy
\begin{equation}\label{eq:phase_constraint}
    h(L-1) \leq 1.
\end{equation}
This constraint restricts the achievable spectral efficiency: for a given modulation index $h$, the maximum alphabet size is $L_{\max} = \lfloor 1/h \rfloor + 1$. 
Violation of~\eqref{eq:phase_constraint} renders coherent demodulation infeasible regardless of the available SNR, as the phase trajectory becomes inherently ambiguous. Consequently, the communication figure of merit $\tilde{\mathcal{C}}$ in Section~\ref{sec:normalized_fom} implicitly assumes that~\eqref{eq:phase_constraint} is satisfied; parameter combinations violating this constraint yield zero achievable rate in practice.

\section{Numerical Results and Discussion}\label{sec:results}

This section presents the practical implementation of the proposed CBS system and validates the analytical framework through numerical evaluation. We first describe the receiver processing chain for separating sensing and communication signals, then discuss a strategy to mitigate interference from clutter or additional targets. Finally, we present numerical results that illustrate the sensing-communication trade-off and system performance.

\subsection{Receiver Processing and Signal Separation}\label{subsec:receiver_processing}

The receiver must extract both range information (sensing) and data symbols (communication) from the beat signal. The processing chain operates as follows. First, the beat signal is sampled at rate $f_s = 1/T_s \geq 2B_{\rm tot}$, yielding the discrete-time representation
\begin{equation}\label{eq:beat_discrete_results}
    d[n] = \mu_s \, e^{j\gamma[n-n_\tau]} \, e^{j(2\pi f_\tau n T_s + \theta_\tau)} + z[n],
\end{equation}
where $n_\tau = \lfloor \tau / T_s \rfloor$ is the discrete delay. The beat frequency $f_\tau = 2\kappa\tau$ encodes the target range, while the phase modulation $\gamma[n]$ carries the communication data.

To separate these components, we employ a data-aided (DA) approach exploiting the preamble structure defined in Section~\ref{subsec:frame_sync}. During the preamble interval, the modulation phase is known, reducing the signal to a tone at frequency $f_\tau$ with a deterministic phase offset. The beat frequency is estimated via maximum likelihood:
\begin{equation}\label{eq:freq_est_ml}
    \hat{f}_\tau = \arg\max_f \left| \sum_{n=0}^{L_p-1} d[n] \, e^{-j(2\pi f n T_s + \gamma_p[n])} \right|^2,
\end{equation}
which is efficiently implemented using the FFT. The range estimate follows as $\hat{R} = c \hat{f}_\tau / (2\kappa)$. Once $\hat{f}_\tau$ is obtained, the beat signal is compensated by multiplying with $e^{-j2\pi\hat{f}_\tau n T_s}$, removing the range-induced frequency shift. The residual signal contains only the communication phase $\gamma[n]$. Data symbols are then recovered via Viterbi decoding, which performs maximum-likelihood sequence estimation by exploiting the CPM trellis structure~\cite{anderson2013digital}, or via a reduced-complexity correlator-bank detector~\cite{proakis2001digital}.

The achievable rate accounting for symbol errors is
\begin{equation}\label{eq:rate_with_ser}
    R = \frac{L_d}{T_{\rm pri}} \left( \log_2 L \right) \left( 1 - P_s \right),
\end{equation}
where $P_s$ denotes the symbol error rate (SER). For the effective SINR $\rho_{\rm eff}$ derived in~\eqref{eq:sinr_xi}, the SER of $L$-ary orthogonal signaling is well approximated by $P_s \approx (L-1)\, Q\bigl(\sqrt{\rho_{\rm eff} \log_2 L}\bigr)$ at moderate-to-high SNR~\cite{proakis2001digital}.

In multi-target scenarios, clutter returns appear as additional tones at positive beat frequencies. Interference mitigation can be achieved by introducing a deliberate Doppler offset at the ST-RIS, which shifts the communication signal to negative frequencies, allowing it to be isolated via filtering. This extension is left for future work.

\subsection{Simulation Parameters}\label{subsec:siL_params}

This section validates the analytical framework developed in Sections~\ref{sec:performance}--\ref{sec:optimization} through numerical evaluation. We first examine the fundamental sensing-communication trade-off captured by the Pareto frontier, then analyze sensing performance under the MCRB, followed by frame synchronization in the presence of residual frequency errors, and finally communication performance including the coupling effects.

The simulation parameters are summarized in Table~\ref{tab:sim_params}.

\begin{table}[b]
    \centering
    \caption{Simulation Parameters}
    \label{tab:sim_params}
    \begin{tabular}{lcc}
        \toprule
        \textbf{Parameter} & \textbf{Symbol} & \textbf{Value} \\
        \midrule
        Carrier frequency & $f_c$ & $77$~GHz \\
        Total bandwidth & $B_{\rm tot}$ & $100$~MHz \\
        Observation time & $T_0$ & $20$~$\mu$s \\
        \midrule
        Modulation & --- & CP-FSK \\
        Modulation index & $h$ & $0.1$ \\
        Alphabet size & $L$ & $4$, $8$ \\
        Preamble length & $L_p$ & $2$, $4$, $8$~symbols \\
        \midrule
        Number of elements & $M$ & $16 \times 16$ \\
        Element spacing & $d$ & $\lambda_c / 4$ \\
        \bottomrule
    \end{tabular}
\end{table}

\subsection{Sensing-Communication Trade-off}\label{subsec:results_tradeoff}

\begin{figure}[t]
    \centering
    \includegraphics[width=0.75\columnwidth]{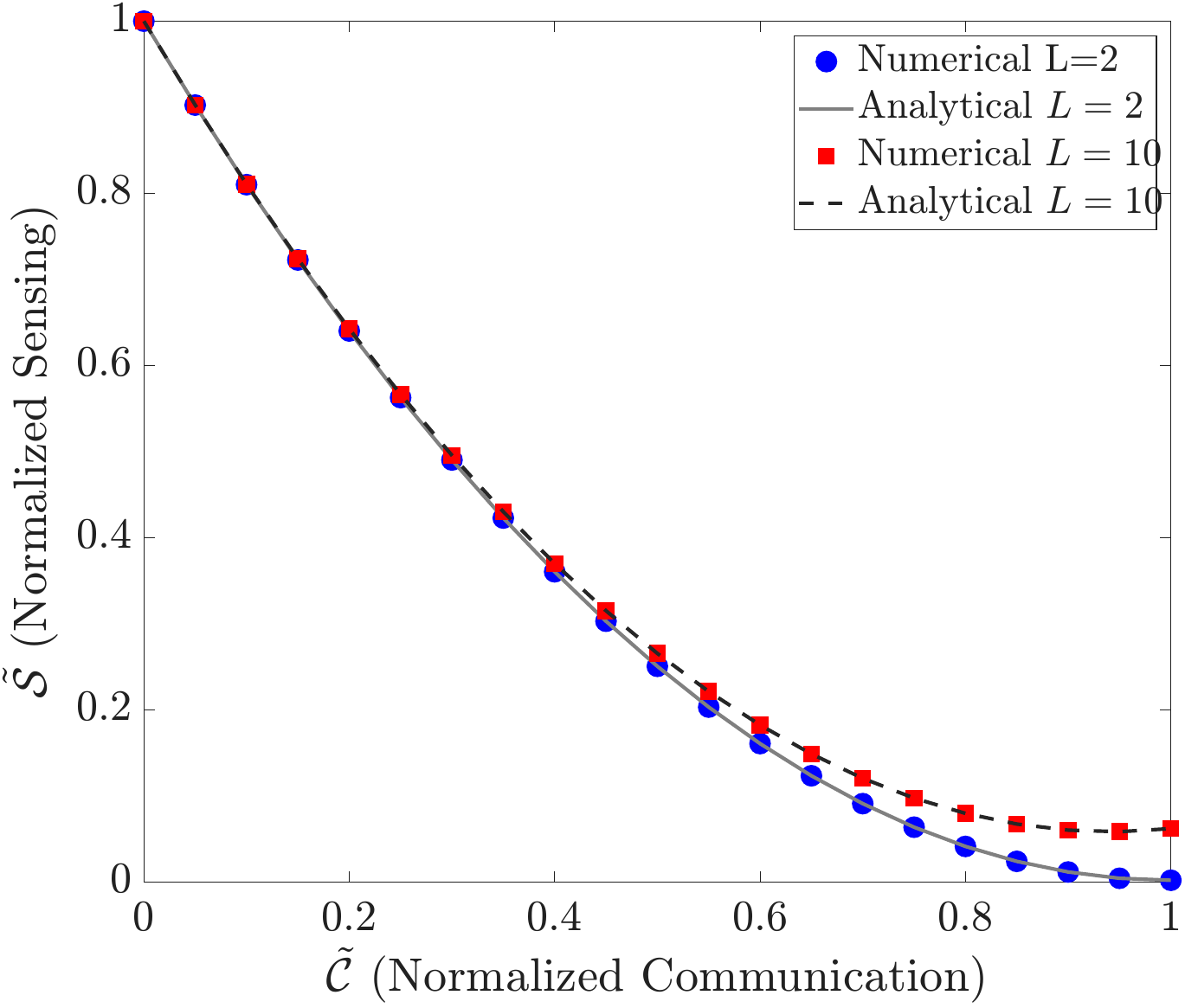}
    \caption{Pareto frontier of the sensing-communication trade-off. Markers denote numerical evaluation; solid lines represent the analytical expression~\eqref{eq:pareto_frontier}. With $h = 0.1$, the ratio $a/b \approx 53$ for $L = 2$ and $a/b \approx 5.3$ for $L = 10$, reflecting the increased modulation contribution to Fisher Information at larger alphabet sizes. Achieving $\tilde{\mathcal{S}} = 0.9$ limits communication to $\tilde{\mathcal{C}} \approx 0.15$.}
    \label{fig:pareto}
\end{figure}

The fundamental trade-off between sensing and communication is governed by the Pareto frontier derived in Section~\ref{sec:normalized_fom}. Fig.~\ref{fig:pareto} presents the normalized sensing figure of merit $\tilde{\mathcal{S}}$ versus the normalized communication figure of merit $\tilde{\mathcal{C}}$ for two alphabet sizes. The analytical expression~\eqref{eq:pareto_frontier}
accurately predicts the simulated trade-off curve. The coefficient ratio $a/b = 16/(3h^2\sigma_b^2)$ quantifies the relative efficiency of the chirp waveform versus the phase modulation in extracting Fisher Information for delay estimation.

Two key observations emerge from Fig.~\ref{fig:pareto}. First, the trade-off is inherently convex: improving communication performance necessarily degrades sensing, with no operating point achieving both $\tilde{\mathcal{S}} = 1$ and $\tilde{\mathcal{C}} = 1$ simultaneously. Second, the alphabet size $L$ influences the trade-off severity through the symbol variance $\sigma_b^2 = (L^2 - 1)/3$. Larger alphabets reduce the ratio $a/b$, yielding a less steep frontier and more favorable trade-off. For $L = 10$, the curve approaches the diagonal, indicating that the modulation contributes substantially to sensing information.

From a design perspective, the Pareto frontier enables systematic resource allocation. Given a minimum sensing requirement $\tilde{\mathcal{S}}_{\min}$, the optimal bandwidth allocation follows from~\eqref{eq:beta_optimal}.

\begin{figure}[t]
    \centering
    \subfloat[MSE vs SNR]{\includegraphics[width=0.75\columnwidth]{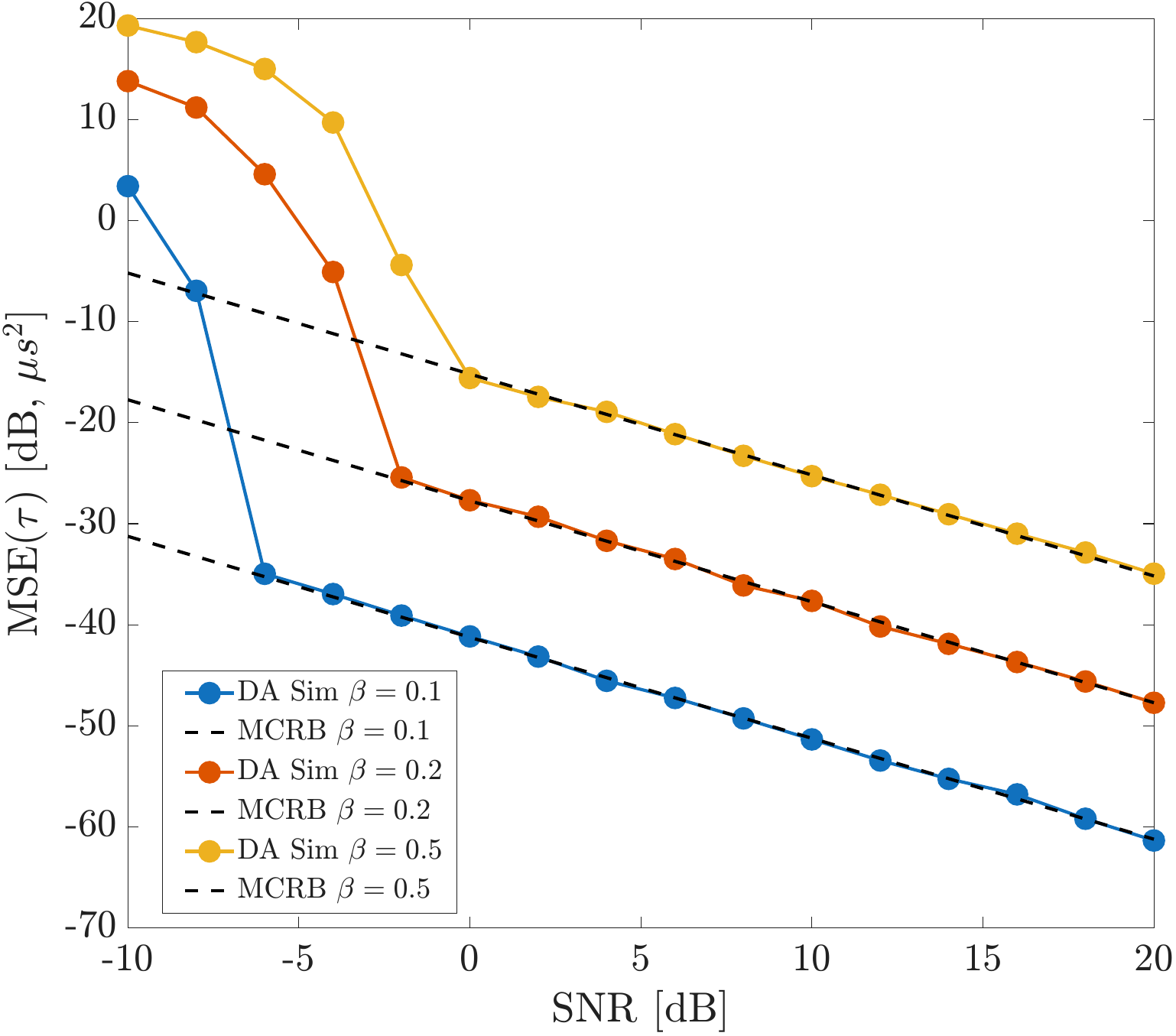}\label{fig:sensing_snr}}
    \\
    \subfloat[MSE vs $\beta$]{\includegraphics[width=0.75\columnwidth]{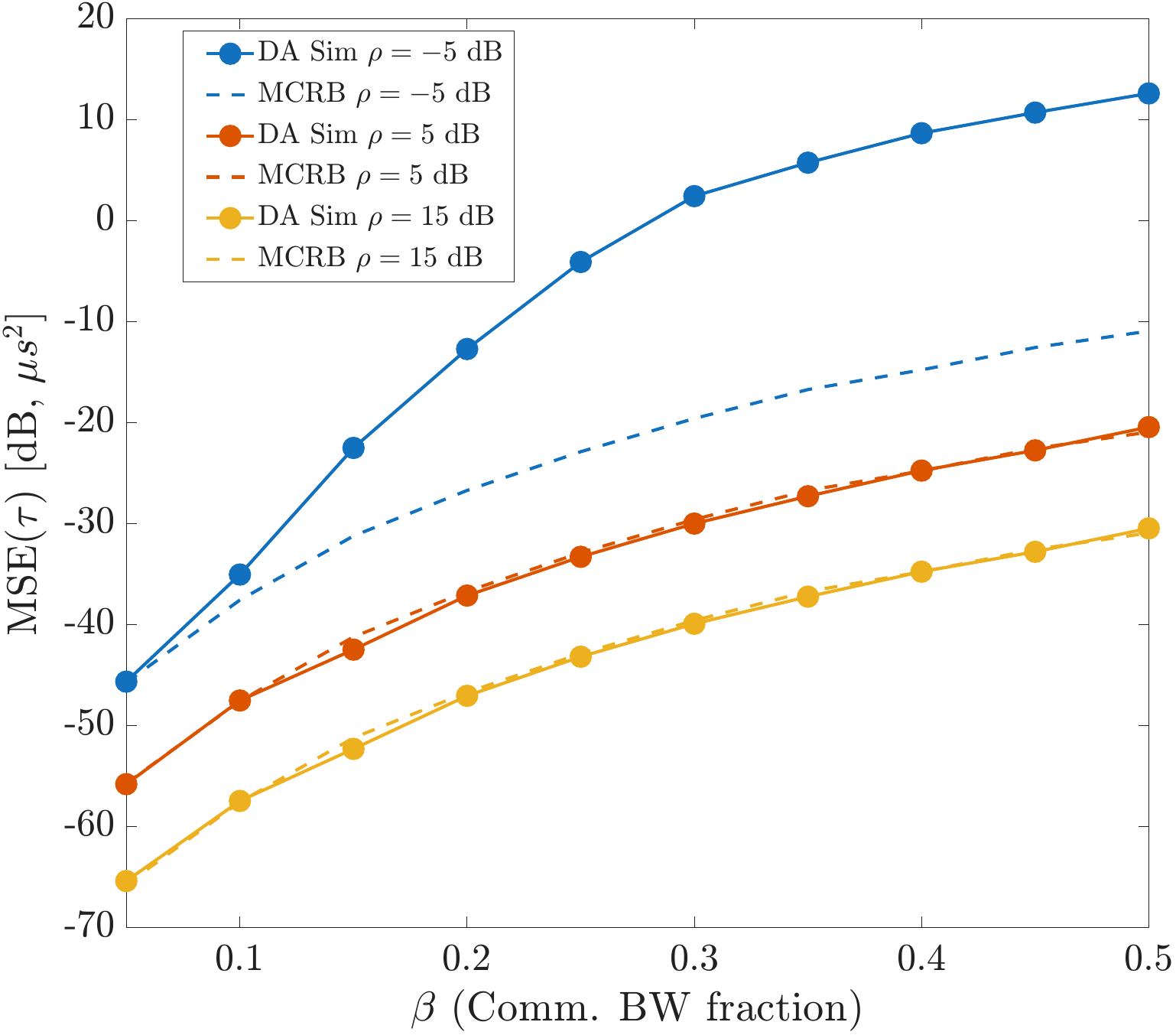}\label{fig:sensing_beta}}
    \caption{Delay estimation performance. (a)~MSE versus SNR for $\beta \in \{0.1, 0.2, 0.5\}$; the DA estimator attains the MCRB (solid lines) across the SNR range. (b)~MSE versus bandwidth allocation at SNR $\in \{-5, 5, 15\}$~dB; at high SNR, larger $\beta$ is tolerable without violating sensing constraints.}
    \label{fig:sensing_performance}
\end{figure}

\subsection{Sensing Performance}\label{subsec:results_sensing}

The MCRB derived in~\eqref{eq:mcrb_final} provides a lower bound on delay estimation accuracy in the presence of unknown data symbols. Fig.~\ref{fig:sensing_performance} presents the sensing performance under different operating conditions. The data-aided (DA) estimator, which exploits the known preamble structure~\cite{damico1994modified}, achieves the MCRB across the evaluated parameter range.

Fig.~\ref{fig:sensing_snr} shows MSE versus SNR for three bandwidth allocations. The DA estimator is statistically efficient: simulated MSE coincides with the theoretical MCRB. Increasing $\beta$ systematically degrades performance, consistent with reduced chirp bandwidth $B_s = (1-\beta)B_{\rm tot}$. Fig.~\ref{fig:sensing_beta} examines MSE versus $\beta$ at fixed SNR values. At low SNR ($\rho = -5$~dB), MSE exhibits steep dependence on $\beta$, reflecting reduced margin for bandwidth reallocation. At high SNR ($\rho = 15$~dB), larger $\beta$ is tolerable while maintaining acceptable accuracy, suggesting an SNR-adaptive allocation strategy.

\subsection{Frame Synchronization}\label{subsec:results_sync}

\begin{figure}[t]
    \centering
    \subfloat[ROC curves]{\includegraphics[width=0.75\columnwidth]{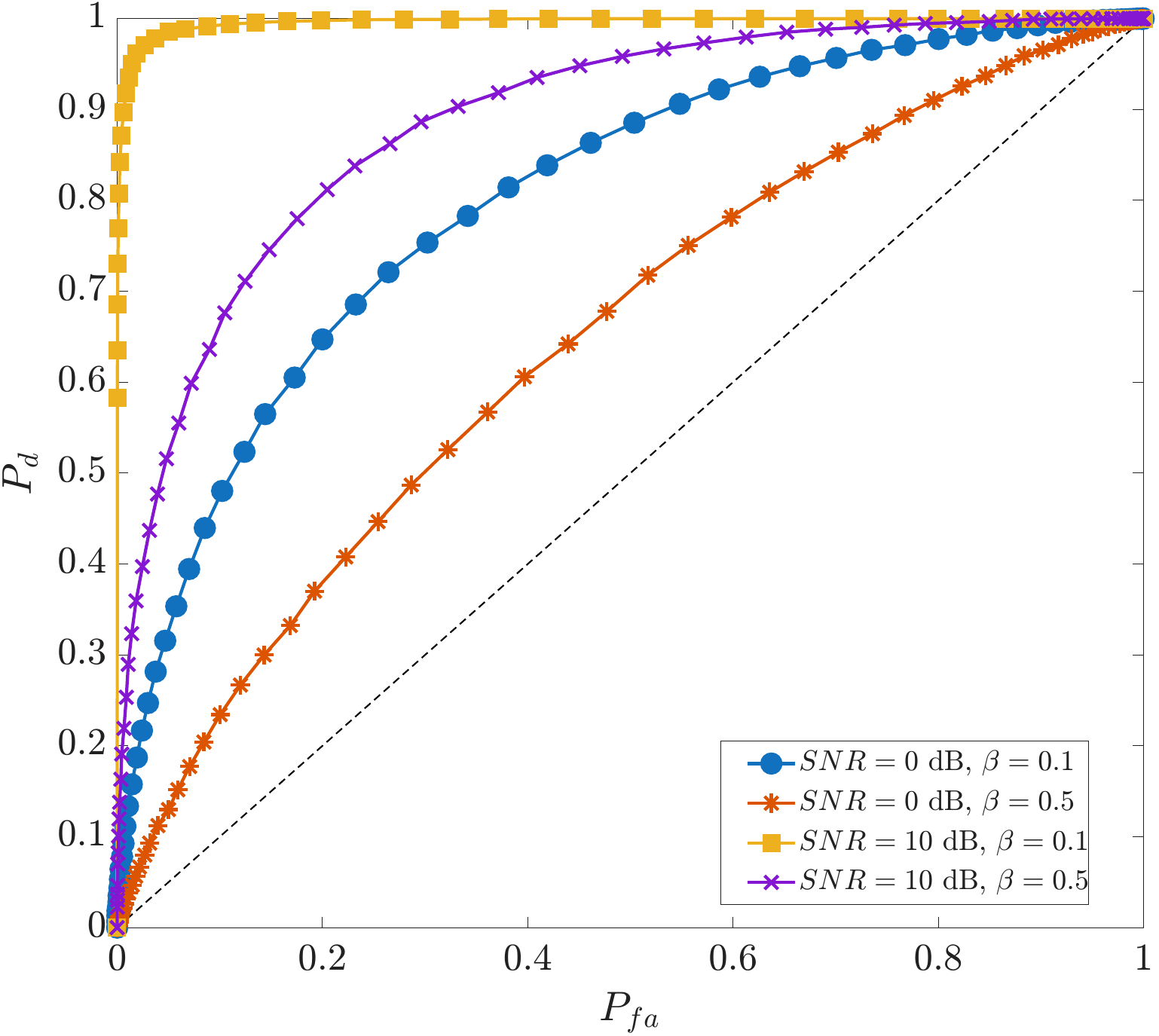}\label{fig:roc}}
    \\
    \subfloat[Detection probability vs SNR]{\includegraphics[width=0.75\columnwidth]{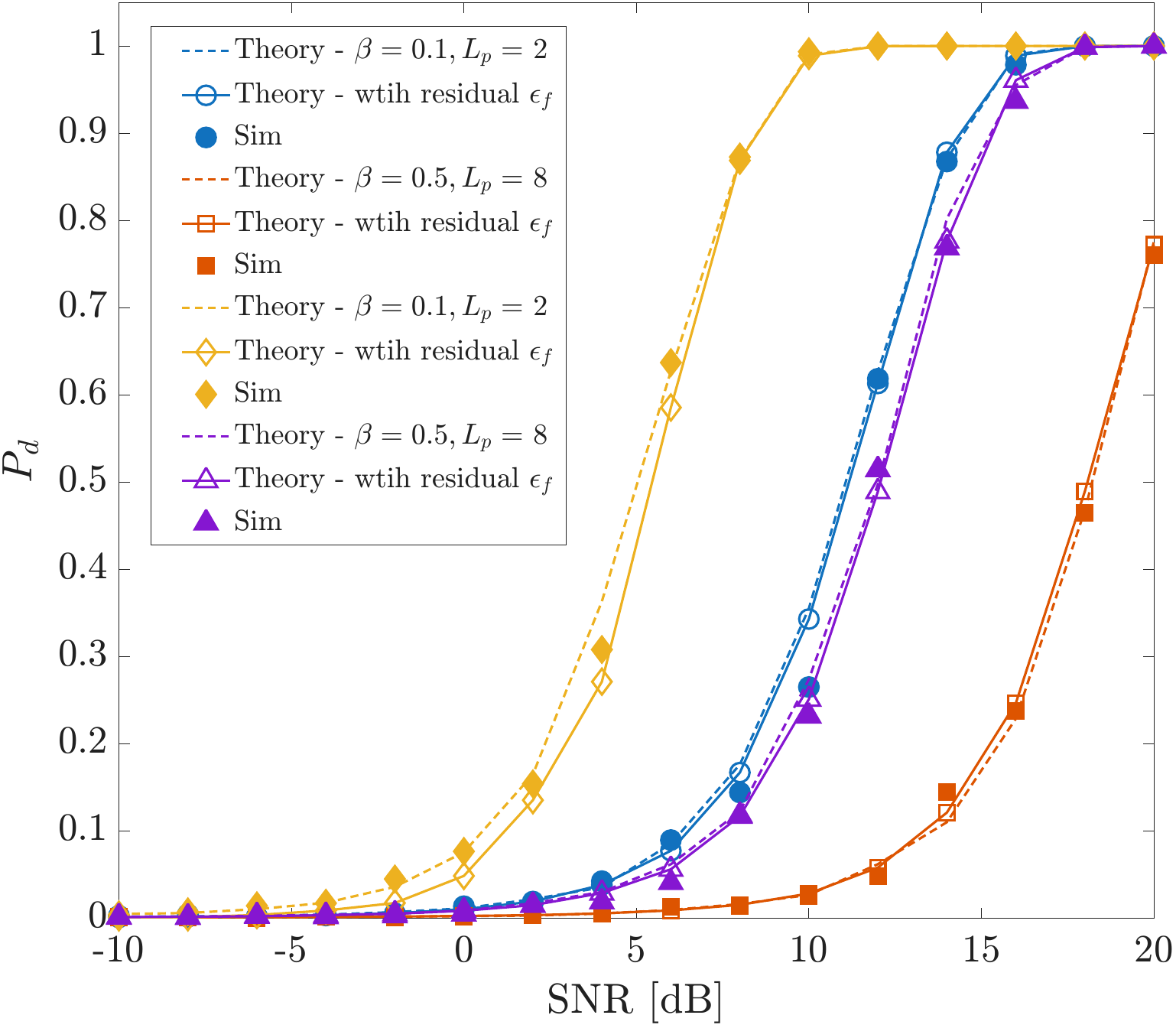}\label{fig:pd_snr}}
    \caption{Frame synchronization performance. (a)~ROC curves for different SNR and bandwidth allocations; higher SNR and lower $\beta$ yield improved detection. (b)~Detection probability at $P_{\rm fa} = 10^{-3}$; dashed lines: theory without residual CFO; solid lines: theory with $\sigma_{\epsilon_f}$ from MCRB; markers: simulation. The gap quantifies the sensing-synchronization coupling.}
    \label{fig:sync_performance}
\end{figure}

Reliable demodulation requires accurate frame timing acquisition. The GLRT detector developed in Section~\ref{subsec:frame_sync} distinguishes between the preamble-modulated signal ($\mathcal{H}_1$) and the unmodulated tone ($\mathcal{H}_0$). A critical consideration is the residual frequency error $\epsilon_f$ arising from imperfect delay estimation, whose variance is bounded by $\sigma_{\epsilon_f}^2 = 4\kappa^2 \cdot \mathrm{MCRB}(\tau)$.

Fig.~\ref{fig:sync_performance} presents the synchronization performance. The ROC curves in Fig.~\ref{fig:roc} show that at $\rho = 10$~dB with $\beta = 0.1$, the detector approaches ideal performance ($P_d > 0.95$ at $P_{\rm fa} = 10^{-2}$), while degradation occurs at lower SNR and higher $\beta$. Fig.~\ref{fig:pd_snr} examines $P_{\rm d}$ versus SNR at $P_{\rm fa} = 10^{-3}$, comparing ideal detection (dashed) with the case incorporating residual CFO from the MCRB (solid). The gap between curves quantifies the sensing-synchronization coupling: at low SNR, where $\sigma_{\epsilon_f}$ is large, residual frequency error significantly degrades detection. As SNR increases, $\sigma_{\epsilon_f} \propto 1/\sqrt{\rho}$ decreases and the curves converge. This behavior has design implications: at low SNR, longer preambles ($L_p = 8$) are necessary, whereas at high SNR, shorter preambles ($L_p = 2$) suffice, maximizing data throughput.
\subsection{Communication Performance}\label{subsec:results_comm}

\begin{figure}[t]
    \centering
    \includegraphics[width=0.75\columnwidth]{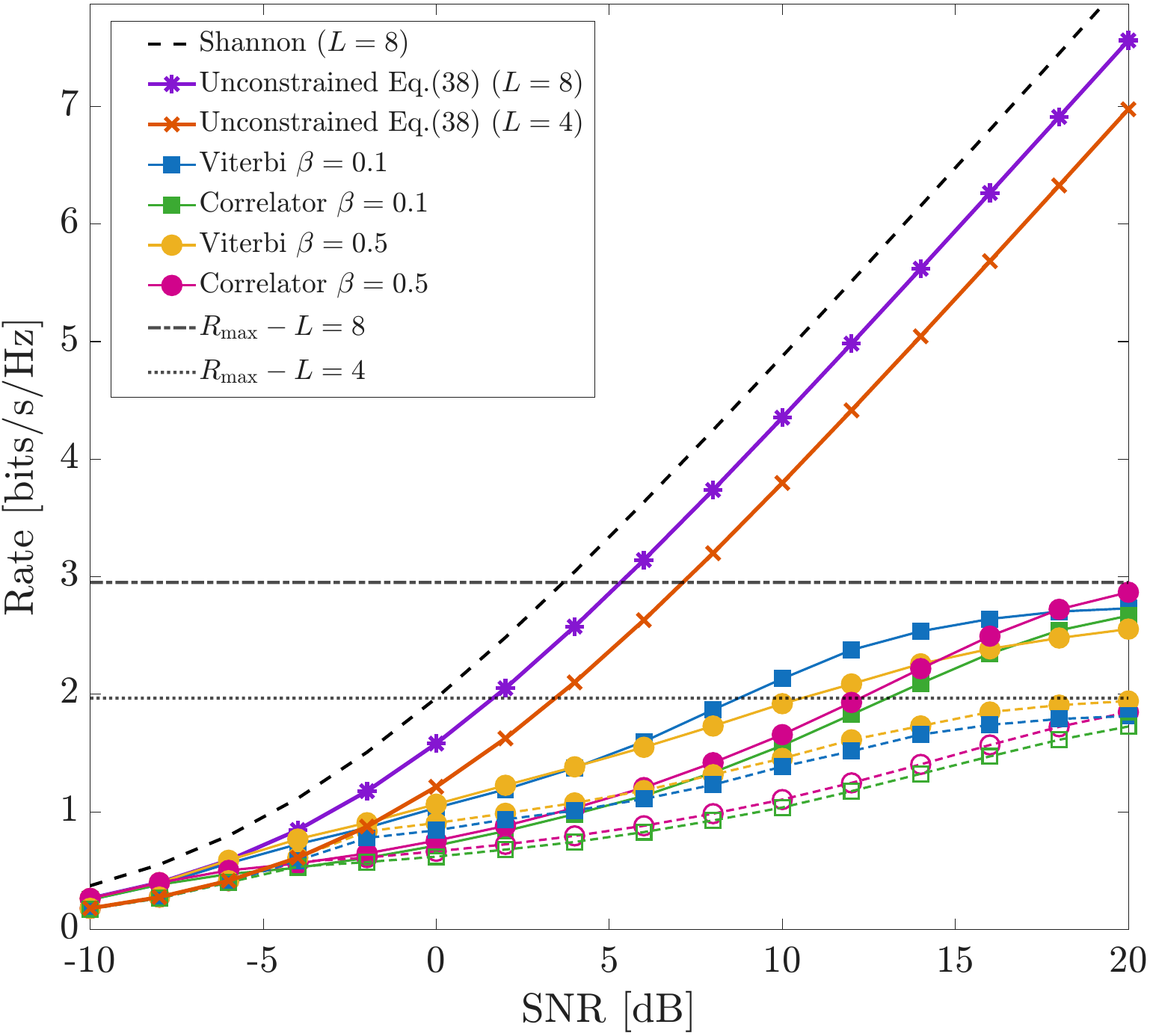}
    \caption{Spectral efficiency versus SNR for $L \in \{4, 8\}$ and $\beta \in \{0.1, 0.5\}$. Dashed lines: Shannon capacity and unconstrained rate from~\eqref{eq:spectral_eff}; dotted lines: maximum rate $R_{\max} = \log_2 L$; markers: Viterbi and correlator detector simulation. At high SNR, rates saturate at $R_{\max}$. The gap from Shannon quantifies the coupling-induced degradation.}
    \label{fig:rate}
\end{figure}

The achievable communication rate depends on the effective SINR after accounting for phase error accumulation, as analyzed in Section~\ref{subsec:comL_perf}. Fig.~\ref{fig:rate} presents the spectral efficiency as a function of SNR for two alphabet sizes ($L = 4$ and $L = 8$) and two bandwidth allocations ($\beta = 0.1$ and $\beta = 0.5$). Several reference curves are included: the Shannon capacity, the unconstrained rate from~\eqref{eq:spectral_eff} accounting for the coupling parameter $\xi$, and the maximum rate $R_{\max} = \log_2 L$ imposed by the finite alphabet.

The results in Fig.~\ref{fig:rate} reveal several important characteristics. First, the unconstrained rate~\eqref{eq:spectral_eff}, which incorporates the coupling parameter $\xi$, accurately predicts the achievable performance in the moderate-SNR regime. The coupling parameter $\xi$ defined in~\eqref{eq:xi_def} captures the phase error accumulation due to residual frequency offset; larger $\xi$ reduces the effective SINR and hence the rate. Second, at high SNR, the rate saturates at $R_{\max} = \log_2 L$, the maximum imposed by the finite modulation alphabet. For $L = 8$, this ceiling is $3$~bits/symbol; for $L = 4$, it is $2$~bits/symbol. Third, the gap between Shannon capacity and the achieved rate quantifies the combined effect of modulation inefficiency and sensing-communication coupling.

Comparing the two detection strategies, Viterbi detection consistently outperforms the correlator bank, particularly at moderate SNR levels, where inter-symbol interference from past symbols becomes significant. The performance gap narrows at high SNR, where both detectors approach error-free operation. From a complexity perspective, the correlator bank offers reduced computational requirements at the cost of approximately $1$--$2$~dB SNR penalty.

The bandwidth allocation $\beta$ influences communication through competing mechanisms. Higher $\beta$ allocates more bandwidth to communication, potentially increasing the symbol rate. However, higher $\beta$ also degrades sensing accuracy, increasing $\sigma_{\epsilon_f}$ and hence the coupling parameter $\xi$. In Fig.~\ref{fig:rate}, the curves for $\beta = 0.1$ and $\beta = 0.5$ exhibit similar performance at low-to-moderate SNR, suggesting that these effects approximately balance. At high SNR, where sensing errors become negligible, the higher symbol rate enabled by $\beta = 0.5$ would yield a throughput advantage (in bits per second), though the spectral efficiency (in bits per second per Hertz) remains comparable.

\section{Conclusions}\label{sec:conclusions}

This paper developed a framework for embedding communication into radar echoes through space-time modulation of a reconfigurable intelligent surface. The transmitted waveform remains a pure sensing signal; information is impressed onto the reflected echo by ST-RIS phase modulation, and the receiver extracts both range and data from the same physical signal.

The MCRB for range estimation under unknown data symbols decomposes into sensing and modulation contributions, both adding constructively to Fisher information. However, under fixed total bandwidth, allocating spectrum to communication reduces the chirp bandwidth, constituting the fundamental sensing-communication trade-off. The Pareto frontier was shown to be convex, and the optimal bandwidth allocation under sensing constraints admits closed-form solution.

Frame synchronization was formulated as a GLRT discriminating preamble-modulated signals from unmodulated tones. Residual frequency error from imperfect range estimation couples sensing performance to synchronization reliability. The achievable rate depends on phase error accumulation captured by the coupling parameter $\xi$: poor range estimation increases frequency error variance, which accumulates as phase drift and reduces effective SNR.

Numerical results validated the analytical framework: the DA estimator attains the MCRB, the GLRT achieves $P_{\rm d} > 0.95$ at moderate SNR, and spectral efficiency predictions align with simulations. The framework applies to scenarios requiring low-power communication with radar terminals, such as passive tagging in automotive radar and sensor reporting in industrial environments. Extensions to multiple ST-RIS devices and mobile scenarios remain for future work.

\appendix
\section{Statistical Analysis of the GLRT}\label{app:glrt_analysis}

This appendix derives the detection and false alarm probabilities for the frame synchronization GLRT introduced in Section~\ref{subsec:frame_sync}.

\subsection{Joint Distribution and Characteristic Function}

The matched-filter outputs $C_p$ and $C_0$ defined in~\eqref{eq:Cp_def}--\eqref{eq:C0_def} share the same noise realization. Under hypothesis $\mathcal{H}_i$, the vector $\mathbf{C} = [C_p, C_0]^{\sf T}$ is jointly complex Gaussian:
\begin{equation}\label{eq:joint_dist}
    \mathbf{C} \sim \mathcal{CN}(\boldsymbol{\nu}_i, \mathbf{R}), \quad
    \mathbf{R} = \sigma^2 \begin{pmatrix} 1 & \nu_c \\ \nu_c^* & 1 \end{pmatrix},
\end{equation}
where $\sigma^2 = L_p \sigma_z^2$ and $\nu_c = \Gamma_p^*/L_p$ is the correlation coefficient arising from integrating identical noise samples. Under perfect frequency estimation, the mean vectors are
\begin{equation}\label{eq:mean_vectors}
    \boldsymbol{\nu}_1 = A e^{j\psi} \begin{pmatrix} L_p \\ \Gamma_p \end{pmatrix}, \quad
    \boldsymbol{\nu}_0 = A e^{j\psi} \begin{pmatrix} \Gamma_p^* \\ L_p \end{pmatrix},
\end{equation}
where $A = |\mu_s|$ and $\Gamma_p = \sum_{n=0}^{L_p-1} e^{j\gamma_p[n]}$ is the preamble DC content.

The GLRT statistic $\Lambda = |C_p|^2 - |C_0|^2$ is a Hermitian quadratic form $\Lambda = \mathbf{C}^{\sf H} \mathbf{Q} \mathbf{C}$ with $\mathbf{Q} = \mathrm{diag}(1, -1)$. For $\mathbf{C} \sim \mathcal{CN}(\boldsymbol{\nu}, \mathbf{R})$, the characteristic function is~\cite{turin1960characteristic}
\begin{equation}\label{eq:cf_lambda}
    \phi_\Lambda(t) = \frac{\exp\left( jt \, \boldsymbol{\nu}^{\sf H} \mathbf{Q} (\mathbf{I} - jt \mathbf{R} \mathbf{Q})^{-1} \boldsymbol{\nu} \right)}{\det(\mathbf{I} - jt \mathbf{R} \mathbf{Q})},
\end{equation}
where the determinant evaluates to $\det(\mathbf{I} - jt \mathbf{R}\mathbf{Q}) = 1 + (\sigma^2 t)^2 (1 - |\nu_c|^2)$. The CDF is obtained via Gil-Pelaez inversion:
\begin{equation}\label{eq:cdf_formula}
    F_\Lambda(\eta) = \frac{1}{2} + \frac{1}{\pi} \int_0^\infty \frac{\mathrm{Im}[\phi_\Lambda(t) e^{-jt\eta}]}{t} \, dt.
\end{equation}
The detection and false alarm probabilities follow as $P_{\rm d} = 1 - F_\Lambda^{(1)}(\eta)$ and $P_{\rm fa} = 1 - F_\Lambda^{(0)}(\eta)$.

\subsection{Effect of Residual Frequency Offset}

Under residual CFO $\epsilon_f \neq 0$, the mean vectors are scaled by the coherence factor $\chi(\epsilon_f) = \sin(\pi \epsilon_f L_p T_s)/\sin(\pi \epsilon_f T_s)$ and modified DC content $\tilde{\Gamma}_p(\epsilon_f)$. The resulting detection probability degrades as
\begin{equation}\label{eq:deflection_cfo}
    P_{\rm d}(\epsilon_f) \approx P_{\rm d}(0) \left( 1 - \frac{(\pi L_p \epsilon_f T_s)^2}{3} \right),
\end{equation}
for $|\epsilon_f| \ll 1/(L_p T_s)$. Performance degrades significantly when the frequency error approaches the coherence bandwidth $1/(L_p T_s)$.

\bibliographystyle{IEEEtran}
\bibliography{Bibliography}

\end{document}